\def\pmb#1{\setbox0=\hbox{#1}%
   \kern-.025em\copy0\kern-\wd0
   \kern.05em\copy0\kern-\wd0
   \kern-0.025em\raise.0433em\box0}
\def\gta{\mathrel{{\lower 3pt\hbox{$\mathchar"218$}}\hskip-8pt
   \raise 2pt\hbox{$\mathchar"13E$}}}
\def\lta{\mathrel{{\lower 3pt\hbox{$\mathchar"218$}}\hskip-8pt
   \raise 2pt\hbox{$\mathchar"13C$}}}
\def\half{{\scriptstyle{1\over2}}}
\def\dagg{\phantom{\dagger}}            

\def\today{\number\day\space\ifcase\month\or
  January\or February\or March\or April\or May\or June\or
  July\or August\or September\or October\or November\or December\fi
 \space\number\year}
\documentstyle[epsfig,twocolumn,aps]{revtex}

\begin{document}
\newcommand{\Vv }{{\raisebox{-1.2pt}{\makebox(0,0){$o$}}}}
\newcommand{\Zz }{{\raisebox{-1.2pt}{\makebox(0,0){$\mbox{\tiny o}$}}}}
\newcommand{\Xx }{{\special{em:moveto}}}
\newcommand{\Yy }{{\special{em:lineto}}}
\newcommand{\Ww }{{\usebox{\plotpoint}}}
\title{\begin{minipage}{6.5in}
{\large \ \\ \ \\ \ \\  \ \\
\centerline{STRIPE-LIKE INHOMOGENEITIES, SPECTROSCOPIES, PAIRING,} 
\centerline{AND COHERENCE IN THE HIGH-$T_c$ CUPRATES} }
{\normalsize  \ \\ 
\centerline{J. ASHKENAZI}
\centerline{Physics Department, University of Miami, P.O. Box 248046,
Coral Gables, FL 33124, U.S.A.} \\ \ \\ \ } 
\end{minipage}
}
\author{\ 
\begin{minipage}{5.125in}
\marginparwidth 0.625in 
\small
\baselineskip 9pt
{\bf Abstract}---It is found that the carriers of the high-$T_c$
cuprates are polaron-like ``stripons'' carrying charge and located in
stripe-like inhomogeneities, ``quasi-electrons'' carrying charge and
spin, and ``svivons'' carrying spin and lattice distortion. This is
shown to result in the observed anomalous spectroscopic properties of
the cuprates. The AF/stripe-like inhomogeneities result from the Bose
condensation of the svivon field, and the speed of their dynamics is
determined by the width of the double-svivon neutron-resonance peak.
Pairing results from transitions between pair states of stripons and
quasi-electrons through the exchange of svivons. The obtained pairing
symmetry is of the $d_{x^2-y^2}$ type; however, sign reversal through
the charged stripes results in features not characteristic of this
symmetry. The phase diagram is determined by a pairing and a coherence
line, associated with a Mott transition, and the pseudogap state
corresponds to incoherent pairing. 
\end{minipage}
}
\maketitle
\setlength{\unitlength}{1in}
\makeatletter
\global\@specialpagefalse
\def\@oddhead{\footnotesize \it \hfill \ Journal of Physics and 
Chemistry of Solids}
\makeatother
\baselineskip 12pt
\normalsize \rm
%
\section{INTRODUCTION}

The existence of stripe-like inhomogeneities in the high-$T_c$ cuprates
has been predicted theoretically quite early \cite{Zaanen}. A wealth of
experimental data support the existence of such inhomogeneities, at
least dynamically, in the superconducting (SC) and the pseudogap states.
The underlying striped structure is characterized by narrow charged
stripes forming antiphase domain walls between wider antiferromagnetic
(AF) stripes \cite{Tran}. 

Experimental observations in the cuprates have been also pointing to the
presence of both itinerant and almost localized (or polaron-like)
carriers in the cuprates. First-principles calculations \cite{Andersen}
support an approach based on the existence of both large-$U$ and
small-$U$ orbitals in the vicinity of the Fermi level ($E_{_{\rm F}}$),
and the applicability of a $t$--$t^{\prime}$--$J$ model for the CuO$_2$
planes.

A small-$U$ electron in band $\nu$, spin $\sigma$ (which is assigned a
number $\pm 1$), and wave vector ${\bf k}$ is created by
$c_{\nu\sigma}^{\dagger}({\bf k})$. The large-$U$ electrons in the
CuO$_2$ planes are approached by the ``slave-fermion'' method
\cite{Barnes}. Such an electron in site $i$ and spin $\sigma$ is created
by $d_{i\sigma}^{\dagger} = e_i^{\dagger} s_{i,-\sigma}^{\dagg}$, if it
is in the ``upper-Hubbard-band'', and by $d_{i\sigma}^{\prime\dagger} =
\sigma s_{i\sigma}^{\dagger} h_i^{\dagg}$, if it is in a Zhang-Rice-type
``lower-Hubbard-band''. Here $e_i^{\dagg}$ and $h_i^{\dagg}$ are
(``excession'' and ``holon'') fermion operators, and
$s_{i\sigma}^{\dagg}$ are (``spinon'') boson operators. These auxiliary
operators have to satisfy the constraint: $e_i^{\dagger} e_i^{\dagg} +
h_i^{\dagger} h_i^{\dagg} + \sum_{\sigma} s_{i\sigma}^{\dagger}
s_{i\sigma}^{\dagg} = 1$. 

An auxiliary space is determined within which a chemical-potential-like
Lagrange multiplier is introduced to impose the constraint on the
average. Physical observables are projected into the physical space by
expressing them as combinations of Green's functions of the auxiliary
space. Since the time evolution of Green's functions is determined by
the Hamiltonian under which the constraint is obeyed rigorously, it is
not expected to be violated as long as justifiable approximations are
used. 

Some of the material presented here, as well as further material about
transport properties, has been published earlier \cite{Ashk1}. 

\section{UNCOUPLED AUXILIARY FIELDS}

The uncoupled spinon field is diagonalized by applying the Bogoliubov
transformation for bosons \cite{Ashk2}: $s_{\sigma}^{\dagg}({\bf k}) =
\cosh{(\xi_{\sigma{\bf k}})} \zeta_{\sigma}^{\dagg}({\bf k}) +
\sinh{(\xi_{\sigma{\bf k}})} \zeta_{-\sigma}^{\dagger}(-{\bf k})$. The
operators $\zeta_{\sigma}^{\dagger}({\bf k})$ create spinon states of
``bare'' energies $\epsilon^{\zeta} ({\bf k})$ which have a V-shape zero
minimum at ${\bf k}={\bf k}_0$. Bose condensation results in AF order at
wave vector ${\bf Q}=2{\bf k}_0 = {\bf Q} = ({\pi \over a} , {\pi \over
a})$. The values of ${\bf k}$ are within the lattice Brillouin zone
(BZ) \cite{Comm1}, within which there are four inequivalent
possibilities for ${\bf k}_0$: $\pm({\pi \over 2a} , {\pi \over 2a})$
and $\pm({\pi \over 2a} , -{\pi \over 2a})$, thus introducing a broken
symmetry. One has $\cosh{(\xi_{{\bf k}})} \cong -\sinh{(\xi_{{\bf k}})}
\gg 1$ for ${\bf k} \to {\bf k}_0$ \cite{Ashk2}. 

The adiabatic approximation is used concerning the dynamics of the
stripe-like inhomogeneities, which are treated as static with respect to
the electrons dynamics. Within the one dimensional charged stripes it is
justified to use the spin-charge separation approximation under which
two-particle spinon-holon (spinon-excession) Green's functions are
decoupled into single-auxiliary-particle Green's functions, resulting in
the existence of a physical interpretation for the auxiliary particles.
Holons (excessions) within the charged stripes are referred to as
``stripons''. Their creation operators (creating charge $-{\rm e}$) are
denoted by $p^{\dagger}_{\mu}({\bf k})$, and bare energies by
$\epsilon^p_{\mu}({\bf k})$. 

Because of the disordered one-dimensional nature of the charged stripes,
it if found appropriate to assume localized uncoupled stripon states.
Their ${\bf k}$ wave vectors present ${\bf k}$-symmetrized combinations
of degenerate localized states to be treated in a perturbation
expansion. They are determined within a BZ based on periodic supercells
which are large enough to approximately contain (each) the entire
spectrum $\epsilon^p_{\mu}$ of bare stripon energies. 

Away from the charged stripes, creation operators of approximate fermion
basis states of coupled holon-spinon  and excession-spinon pairs are
constructed, within the auxiliary space, as follows: 
\begin{eqnarray}
f_{\lambda\sigma}^{\dagger}({\bf k}^{\prime}, {\bf k}) &=&
{e_{\lambda}^{\dagger}({\bf k}^{\prime})
s_{\lambda,-\sigma}^{\dagg}({\bf k}^{\prime} - {\bf k}) \over
\sqrt{n^e_{\lambda}({\bf k}^{\prime}) + n^s_{\lambda,-\sigma}({\bf
k}^{\prime} - {\bf k})}}, \\ 
g_{\lambda\sigma}^{\dagger}({\bf k}^{\prime}, {\bf k}) &=& {\sigma
h_{\lambda}^{\dagg}({\bf k}^{\prime}) s_{\lambda\sigma}^{\dagger}({\bf
k} - {\bf k}^{\prime}) \over \sqrt{n^h_{\lambda}({\bf k}^{\prime}) +
n^s_{\lambda\sigma}({\bf k} - {\bf k}^{\prime})}}, 
\end{eqnarray} 
where ${\bf k}$ and ${\bf k}^{\prime}$ are within the lattice BZ, the
index $\lambda$ accounts for the structure of the unit cell, and:
$n^e_{\lambda}({\bf k}) \equiv \langle e_{\lambda}^{\dagger}({\bf k})
e_{\lambda}^{\dagg}({\bf k}) \rangle$, $n^h_{\lambda}({\bf k}) \equiv
\langle h_{\lambda}^{\dagger}({\bf k}) h_{\lambda}^{\dagg}({\bf k})
\rangle$, $n^s_{\lambda\sigma}({\bf k}) \equiv \langle
s_{\lambda\sigma}^{\dagger}({\bf k}) s_{\lambda\sigma}^{\dagg}({\bf k})
\rangle$. 

The states created by $f_{\lambda\sigma}^{\dagger}({\bf k}^{\prime},
{\bf k})$ and $g_{\lambda\sigma}^{\dagger}({\bf k}^{\prime}, {\bf k})$
have to be orthogonalized to the stripon states, and depleted to avoid
over-completeness. Together with the small-$U$ states [created by
$c_{\nu\sigma}^{\dagger}({\bf k})$] they form, within the auxiliary
space, a basis to ``quasi-electron'' (QE) states whose creation
operators are expressed as combinations:
\begin{eqnarray}
q_{\iota\sigma}^{\dagger}({\bf k}) &=& \sum_{\nu} U^{cq}_{\nu
\iota}(\sigma{\bf k})^* c_{\nu\sigma}^{\dagger}({\bf k}) + \sum_{\lambda
{\bf k}^{\prime}} \big[ U^{fq}_{\lambda\iota}(\sigma{\bf k}^{\prime},
\sigma{\bf k})^* \nonumber \\ 
&\times& f_{\lambda\sigma}^{\dagger}({\bf k}^{\prime}, {\bf k}) +
U^{gq}_{\lambda\iota}(\sigma{\bf k}^{\prime}, \sigma{\bf k})^*
g_{\lambda\sigma}^{\dagger}({\bf k}^{\prime}, {\bf k}) \big], 
\end{eqnarray} 
where the $U$ coefficients are the eigenvector elements obtained in the
diagonalization of the Hamiltonian within a mean-field approximation
(and no coupling to the stripon and spinon fields). The obtained
eigenvalues are the bare QE energies $\epsilon^q_{\iota} ({\bf k})$
which form quasi-continuous ranges of bands within the BZ around
$E_{_{\rm F}}$ \cite{Comm2}. 

\section{COUPLED AUXILIARY FIELDS}

Hopping and hybridization terms introduce strong coupling between the
QE, stripon and spinon fields. It can be expressed in terms of a
coupling Hamiltonian whose parameters can be in principle derived
self-consistently from the original Hamiltonian. For p-type cuprates
this coupling Hamiltonian can be expressed as: 
\begin{eqnarray}
{\cal H}^{\prime} &=& {1 \over \sqrt{N}} \sum_{\iota\lambda\mu\sigma}
\sum_{{\bf k}, {\bf k}^{\prime}} \big\{\sigma
\epsilon^{qp}_{\iota\lambda\mu}(\sigma{\bf k}, \sigma{\bf k}^{\prime})
q_{\iota\sigma}^{\dagger}({\bf k}) p_{\mu}^{\dagg}({\bf k}^{\prime})
\nonumber \\ &\ &\times [\cosh{(\xi_{\lambda,\sigma({\bf k} - {\bf
k}^{\prime})})} \zeta_{\lambda\sigma}^{\dagg}({\bf k} - {\bf
k}^{\prime}) \nonumber \\ &\ &+ \sinh{(\xi_{\lambda,\sigma({\bf k} -
{\bf k}^{\prime})})} \zeta_{\lambda,-\sigma}^{\dagger}({\bf k}^{\prime}
- {\bf k})] + h.c. \big\}. 
\end{eqnarray} 
Here ${\bf k}$ is within the stripons BZ, where the lattice BZ has been
embedded, redefining the band indices of the QE's and the spinons
appropriately. Using the Green's functions formalism, the QE, stripon
and spinon propagators are couples by a vertex introduced through ${\cal
H}^{\prime}$ \cite{Ashk3}. 

The stripe-like inhomogeneities involve the lattice \cite{Bian} due to
its coupling to the stripons. Consequently, processes (induced by ${\cal
H}^{\prime}$) involving a transitions between stripon and QE states,
followed by the emission and/or absorption of spinons, should involve
also the emission and/or absorption of phonons. This can be expressed by
multiplying a spinon propagator, in such processes, by a power series of
phonon propagators \cite{Ashk3}. Such a phonon-``dressed'' spinon is
referred to as a ``svivon'', carrying spin and lattice distortion, and
it replaces the spinon in the ${\cal H}^{\prime}$ vertex. 

\begin{figure}[b!] 
\centerline{\epsfig{file=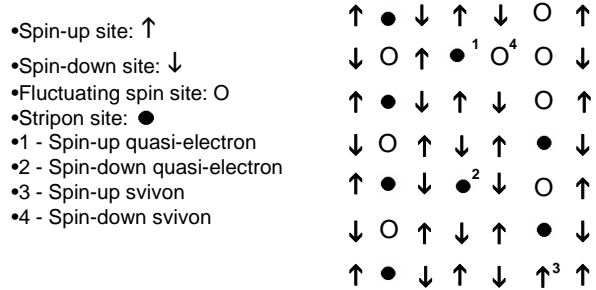,width=3.25in}}
\vspace{10pt}
\caption{An adiabatic ``snapshot'' of a stripe-like inhomogeneity and 
carriers within a CuO$_2$ plane.} 
\label{fig1}
\end{figure}

An adiabatic ``snapshot'' of a CuO$_2$ plane with a stripe-like
inhomogeneity, and physical realizations of the auxiliary fields in the
plane, within the $t$--$t^{\prime}$--$J$ model, is shown in Fig.~1.
Within the adiabatic time scale a site is ``spinless'' either if the
spin on it is replaced by charge, removing the electron or the hole
carrying that spin (as in ``stripon sites'' in Fig.~1), or if the spin
is fluctuating on a shorter time scale (due to, {\it e.g.}, being in
fluctuating singlet spin pairs). A site stripon excitation represents a
transition between these two types of a spinless site within the charged
stripes. A site svivon excitation represents a transition between a
spinned site and a fluctuating-spin spinless site. A site QE excitation
represents a transition between a spinned site and a charged spinless
site within the AF stripes. For simplicity we ignore in Fig.~1 the
dynamics of the QE, stripon, and svivon sites (whose time scale is
shorter than the adiabatic time scale). 

The Green's functions ${\cal G}$ of the auxiliary fields are approached
through the spectral functions $A^q_{\iota}({\bf k}, \omega)$,
$A^p_{\mu}({\bf k}, \omega)$, and $A^{\zeta}_{\lambda}({\bf k}, \omega)$
for the QE's, stripons, and svivons, respectively [$A(\omega) \equiv \Im
{\cal G}(\omega-i0^+) / \pi$]. The solution studied (see below) has a
considerably smaller stripon bandwidth than the QE and svivon
bandwidths; thus a phase-space argument can be used, as in the Migdal
theorem, to prove that vertex corrections to the ${\cal H}^{\prime}$
vertex are negligible. The resulting diagrams \cite{Ashk3} for the
self-energy $\Sigma$ yield the following expressions for the QE,
stripon, and svivon scattering rates $\Gamma^q({\bf k}, \omega)$,
$\Gamma^p({\bf k}, \omega)$, and $\Gamma^{\zeta}({\bf k}, \omega)$
[$\Gamma({\bf k}, \omega) \equiv 2\Im \Sigma({\bf k}, \omega-i0^+)$]: 
\begin{eqnarray}
\Gamma&^q_{\iota\iota^{\prime}}&({\bf k}, \omega) \cong {2\pi \over N}
\sum_{\lambda\mu{\bf k}^{\prime}} \int d\omega^{\prime}
\epsilon^{qp}_{\iota\lambda\mu}({\bf k}^{\prime}, {\bf
k})\epsilon^{qp}_{\iota^{\prime}\lambda\mu}({\bf k}^{\prime}, {\bf k})^*
\nonumber \\ &\times& A^p_{\mu}({\bf k}^{\prime}, \omega^{\prime})
[-\cosh{^2(\xi_{\lambda,{\bf k} - {\bf k}^{\prime}})}
A^{\zeta}_{\lambda}({\bf k} - {\bf k}^{\prime}, \omega -
\omega^{\prime}) \nonumber \\ &+& \sinh{^2(\xi_{\lambda,{\bf k} - {\bf
k}^{\prime}})} A^{\zeta}_{\lambda}({\bf k} - {\bf k}^{\prime},
\omega^{\prime} - \omega)] \nonumber \\ &\times&
[f_{_T}(\omega^{\prime}) + b_{_T}(\omega^{\prime} - \omega)], \\ 
\Gamma&^p_{\mu\mu^{\prime}}&({\bf k}, \omega) \cong {2\pi \over N}
\sum_{\iota{\bf k}^{\prime}\sigma} \int d\omega^{\prime}
\epsilon^{qp}_{\iota\lambda\mu}({\bf k}^{\prime}, {\bf k})^*
\epsilon^{qp}_{\iota\lambda\mu^{\prime}}({\bf k}^{\prime}, {\bf k})
\nonumber \\ &\times& A^q_{\iota}({\bf k}^{\prime}, \omega^{\prime})
[\cosh{^2(\xi_{\lambda,{\bf k}^{\prime} - {\bf k}})}
A^{\zeta}_{\lambda}({\bf k}^{\prime} - {\bf k}, \omega^{\prime} -
\omega) \nonumber \\ &-& \sinh{^2(\xi_{\lambda,{\bf k}^{\prime} - {\bf
k}})} A^{\zeta}_{\lambda}({\bf k}^{\prime} - {\bf k}, \omega -
\omega^{\prime})] \nonumber \\ &\times& [f_{_T}(\omega^{\prime}) +
b_{_T}(\omega^{\prime} - \omega)], \\ 
\Gamma&^{\zeta}_{\lambda\lambda^{\prime}}&({\bf k}, \omega) \cong {2\pi
\over N} \sum_{\iota{\bf k}^{\prime}\mu} \int d\omega^{\prime}
\epsilon^{qp}_{\iota\lambda\mu}({\bf k}^{\prime}, {\bf k}^{\prime} -
{\bf k})^* \nonumber \\ &\times&
\epsilon^{qp}_{\iota\lambda^{\prime}\mu}({\bf k}^{\prime}, {\bf
k}^{\prime} - {\bf k}) [\cosh{(\xi_{\lambda{\bf k}})}
\cosh{(\xi_{\lambda^{\prime}{\bf k}})} A^q_{\iota}({\bf k}^{\prime},
\omega^{\prime}) \nonumber \\ &\times& A^p_{\mu}({\bf k}^{\prime} - {\bf
k}, \omega^{\prime} - \omega) + \sinh{(\xi_{\lambda{\bf k}})}
\sinh{(\xi_{\lambda^{\prime}{\bf k}})} \nonumber \\ &\times&
A^q_{\iota}({\bf k}^{\prime}, -\omega^{\prime}) A^p_{\mu}({\bf
k}^{\prime} - {\bf k}, \omega - \omega^{\prime})] \nonumber \\ &\times&
[f_{_T}(\omega^{\prime} - \omega)- f_{_T}(\omega^{\prime})], 
\end{eqnarray} 
where $f_{_T}(\omega)$ and $b_{_T}(\omega)$ are the Fermi and Bose
distribution functions at temperature $T$. 

The real parts of the self energies and the spectral functions (based on
the diagonal self-energy terms) are obtained from the above expressions
through: 
\begin{eqnarray}
\Re\Sigma({\bf k}, \omega) &=& \wp\int {d\omega^{\prime} \Gamma({\bf k},
\omega^{\prime}) \over 2\pi (\omega - \omega^{\prime})}, \\ 
A({\bf k}, \omega) &=& {\Gamma({\bf k}, \omega)/2\pi \over [\omega -
\epsilon({\bf k}) - \Re\Sigma({\bf k}, \omega)]^2 + [\Gamma({\bf k},
\omega)/2]^2}. 
\end{eqnarray} 
 
\section{AUXILIARY SPECTRAL FUNCTIONS}

A self-consistent solution is obtained, and expressions are derived
below for the intermediary energy range. The high energy range ($\gta
0.5\;$eV), determined by the hopping and exchange parameters, is treated
by introducing cut-off integration limits at $\pm\omega_c$ to the
integrals (resulting in spurious logarithmic divergencies at
$\pm\omega_c$). The solution includes a low energy range ($\lta
0.02\;$eV), appearing in these expressions as ``zero-energy''
non-analytic behavior. Ignoring the variation of the matrix elements in
Eqs.~(5--7), and omitting the dependence on ${\bf k}$ for simplicity,
the sums of the auxiliary spectral functions over the band indices and
small ${\bf k}^{\prime}$ ranges can expressed, self-consistently, as
(all the coefficients are positive): 
\begin{eqnarray}
\tilde A^q(\omega) &\cong& \cases{a^q_+\omega + b^q_+ \;,& for
$\omega>0$,\cr -a^q_-\omega + b^q_- \;,& for $\omega<0$,\cr} \\ 
\tilde A^p(\omega) &\cong& \delta(\omega), \\ 
\tilde A^{\zeta}(\omega) &\cong& \cases{a^{\zeta}_+\omega + b^{\zeta}_+
\;,& for $\omega>0$,\cr a^{\zeta}_-\omega - b^{\zeta}_- \;,& for
$\omega<0$.\cr} 
\end{eqnarray} 
Analyticity is expected to be restored in the low-energy range, and
specifically $\tilde A^{\zeta}(\omega=0) = 0$. Also special behavior is
expected for svivons around ${\bf k}_0$. By Inserting these terms in
Eqs.~(5--7), the following expressions are derived, assuming the low $T$
limits for $f_{_T}(\omega)$ and $b_{_T}(\omega)$ (again all the
coefficients are positive): 
\begin{eqnarray}
{\Gamma^q(\omega) \over 2\pi} &\cong& \cases{c^q_+\omega + d^q_+ \;,&
for $\omega>0$,\cr -c^q_-\omega + d^q_- \;,& for $\omega<0$,\cr} \\ 
{\Gamma^p(\omega) \over 2\pi} &\cong& \cases{c^p_+\omega^3 +
d^p_+\omega^2 + e^p_+\omega \;,& for $\omega>0$,\cr -c^p_-\omega^3 +
d^p_-\omega^2 - e^p_-\omega \;,& for $\omega<0$,\cr} \\ 
{\Gamma^{\zeta}(\omega) \over 2\pi} &\cong& \cases{c^{\zeta}_+\omega +
d^{\zeta}_+ \;,& for $\omega>0$,\cr c^{\zeta}_-\omega - d^{\zeta}_- \;,&
for $\omega<0$.\cr} 
\end{eqnarray} 
Integrating through Eq.~(8) (between the limits $\pm\omega_c$) results
in: 
\begin{eqnarray}
-\Re\Sigma^q&(\omega)& \cong \omega_c(c^q_+ - c^q_-) + \big(d^q_+
\ln{\big|{\omega - \omega_c \over \omega}\big|} \nonumber \\ &-& d^q_-
\ln{\big|{\omega + \omega_c \over \omega}\big|}\big) + \omega\big(c^q_+
\ln{\big|{\omega - \omega_c \over \omega}\big|} \nonumber \\ &+& c^q_-
\ln{\big|{\omega + \omega_c \over \omega}\big|}\big), \\ 
-\Re\Sigma^p&(\omega)& \cong \big[{\omega_c^3 \over 3}(c^p_+ - c^p_-) +
{\omega_c^2 \over 2}(d^p_+ - d^p_-) \nonumber \\ &+& \omega_c(e^p_+ -
e^p_-)\big] + \omega \big[{\omega_c^2 \over 2}(c^p_+ + c^p_-)  \nonumber
\\ &+& \omega_c(d^p_+ + d^p_-) + e^p_+ \ln{\big|{\omega - \omega_c \over
\omega}\big|} \nonumber \\ &+& e^p_- \ln{\big|{\omega + \omega_c \over
\omega}\big|}\big] + \omega^2 \big[\omega_c(c^p_+ - c^p_-) \nonumber \\
&+& d^p_+ \ln{\big|{\omega - \omega_c \over \omega}\big|} - d^p_-
\ln{\big|{\omega + \omega_c \over \omega}\big|}\big] \nonumber \\ &+&
\omega^3 \big[c^p_+ \ln{\big|{\omega - \omega_c \over \omega}\big|} +
c^p_- \ln{\big|{\omega + \omega_c \over \omega}\big|}\big], \\ 
-\Re\Sigma^{\zeta}&(\omega)& \cong \omega_c(c^{\zeta}_+ + c^{\zeta}_-) +
\big(d^{\zeta}_+ \ln{\big|{\omega - \omega_c \over \omega}\big|} 
\nonumber \\ &+& d^{\zeta}_- \ln{\big|{\omega + \omega_c \over
\omega}\big|}\big) + \omega\big(c^{\zeta}_+ \ln{\big|{\omega - \omega_c
\over \omega}\big|} \nonumber \\ &-& c^{\zeta}_- \ln{\big|{\omega +
\omega_c \over \omega}\big|}\big). 
\end{eqnarray} 
Note that the logarithmic divergencies at $\omega=0$ are truncated by
analyticity in the low-energy range. 

The resulting $\Re\Sigma$ renormalize the auxiliary-particle energies
$\epsilon$ to: $\bar\epsilon = \epsilon + \Re\Sigma(\bar\epsilon)$. The
extent of the renormalization can be estimated through $d \bar\epsilon /
d \epsilon = [1 - d \Re\Sigma(\bar\epsilon) / d \bar\epsilon]^{-1}$.
This renormalization is particularly strong for the stripon energies,
due to the effect of the quasi-continuum of QE bands, reflected in a
significant $\omega^3$ term in $\Gamma^p(\omega)$ (14). It introduces a
large negative term $-(c^p_+ + c^p_-) \omega_c^2/2$ (17) to $d
\Re\Sigma^p(\bar\epsilon) / d \bar\epsilon$, resulting in a very small
$d \bar\epsilon^p / d \epsilon^p$. Thus the stripon bandwidth drops down
to the low energy range, and a $\delta$-function is appropriate for
$\tilde A^p(\omega)$ (11). 

Expressions (10), for $\tilde A^q(\omega)$, and (12) for $\tilde
A^{\zeta}(\omega)$ result, through Eq. (9), from the effects of bands
crossing zero energy, which mainly contribute to the constant ($b$)
terms [due to the normalization of the spectral functions (9)], and of
higher energy bands, whose effect through (9) is approximately $\propto
\Gamma(\omega)$, and thus contribute to the constant ($b$) and linear
($a$) terms. 

The self-consistent treatment introduces inequalities between positive
and negative $\omega$ coefficients in Eqs.~(10--15), resulting from the
fact that the svivon spectrum has more weight for $\omega>0$, and that
[in Eqs.~(5--7)] $\cosh{^2(\xi_{{\bf k}})} > \sinh{^2(\xi_{{\bf k}})}$.
For the case of p-type cuprates (worked out in these equation) the
following inequalities are built up self-consistently: 
\begin{eqnarray}
a^q_+&>&a^q_-,\ \ \  b^q_+>b^q_-,\ \ \  c^q_+>c^q_-,\ \ \  d^q_+>d^q_-,
\\ a^{\zeta}_+&>&a^{\zeta}_-,\ \ \  b^{\zeta}_+>b^{\zeta}_-,\ \ \
c^{\zeta}_+>c^{\zeta}_-,\ \ \  d^{\zeta}_+>d^{\zeta}_-. 
\end{eqnarray} 
For ``real'' n-type cuprates, in which the stripons are based on
excession and not holon states, the roles of $\cosh{(\xi_{{\bf k}})}$
and $\sinh{(\xi_{{\bf k}})}$ are reversed in ${\cal H}^{\prime}$ (4),
and in the expressions derived from it. consequently the direction of
the inequalities is reversed for the QE coefficients (19), but stays the
same for the svivon coefficients (20). One could expect deviations
from the inequalities (19--20) at specific ${\bf k}$ points; they almost
disappear for svivons close to point ${\bf k}_0$, where
$\cosh{(\xi_{{\bf k}})} \cong -\sinh{(\xi_{{\bf k}})}$ \cite{Ashk2}, and
for QE's coupled mainly to such svivons. 

\begin{figure}[b!] 
\centerline{\epsfig{file=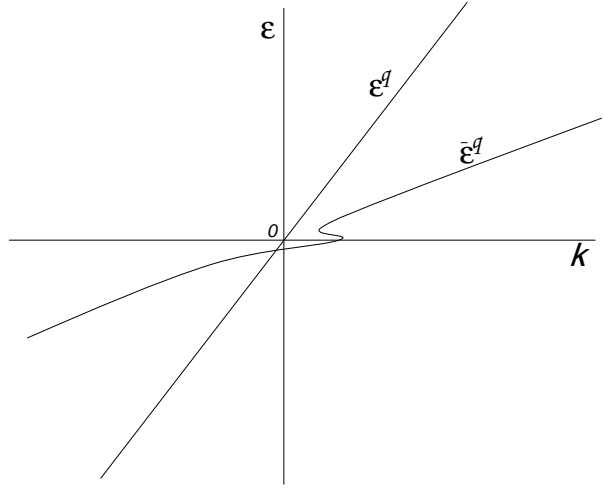,width=3.25in}}
\vspace{10pt}
\caption{A typical self-energy renormalization of the QE energies, for
p-type cuprates.} 
\label{fig2}
\end{figure}

A typical renormalization of the QE energies (for p-type cuprates), 
around zero energy, is shown in Fig.~2. One renormalization effect is a
reduction of the band slope. Another effect is a consequence of the
logarithmic singularity (truncated in the low-energy range) in
$\Re\Sigma^q$ at $\omega=0$, due to the $(d^q_+ - d^q_-)\ln{|\omega|}$
term in Eq.~(16). The asymmetry between positive and negative energies is
a consequence of inequality (19), and this asymmetry is expected to be
inverted for ``real'' n-type cuprates. 

\begin{figure}[b!] 
\centerline{\epsfig{file=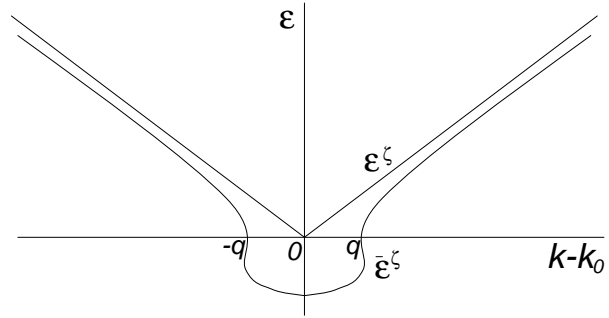,width=3.25in}}
\vspace{10pt}
\caption{A typical self-energy renormalization of the svivon energies
around the minimum at ${\bf k}_0$.} 
\label{fig3}
\end{figure}

A typical renormalization of the svivon energies, around the V-shape
zero minimum of $\epsilon^{\zeta}$ at ${\bf k}_0$, is shown in Fig.~3. A
major renormalization effect is due to the $(d^{\zeta}_+ +
d^{\zeta}_-)\ln{|\omega|}$ term in $\Re\Sigma^{\zeta}(\omega)$ (18),
contributing a logarithmic singularity (truncated in the low-energy
range) at $\omega=0$. By Eq.~(15) $\bar\epsilon^{\zeta}$ is expected to
have a considerable linewidth around ${\bf k}_0$ (except where it
crosses zero). This changes when a pairing gap in low-energy QE and
stripon states (coupled by svivons around ${\bf k}_0$) disables the
scattering processes causing linewidth (7) near the negative minimum of
$\bar\epsilon^{\zeta}$ at ${\bf k}_0$. 

This renormalization of the svivon energies changes the physical
signature of their Bose condensation from AF order to the observed
stripe-like inhomogeneities. The exact low-energy details around the
minimum in $\bar\epsilon^{\zeta}$ determine, self-consistently, the
nature of these inhomogeneities. The dynamics of these inhomogeneities
depends on the linewidth of $\bar\epsilon^{\zeta}({\bf k}_0)$ around
${\bf k}_0$, and thus becomes slower in pairing states (where they can 
be observed). The crossover energy from the intermediary to the low
energy range is determined by $|\bar\epsilon^{\zeta}({\bf k}_0)|$ and
the stripon bandwidth $\omega^p$. 

\section{ELECTRON SPECTRUM}

Spectroscopies, like ARPES, measuring the effect of transfer of
electrons into, or out of, the crystal, are determined by $A_e({\bf p},
\omega) \equiv \Im {\cal G}_e({\bf p}, \omega-i0^+) / \pi$, the
electrons spectral function at momentum ${\bf p}$ and energy $\omega$.
They are expressed in terms of the auxiliary space spectral functions in
Eqs.~(10--12), thus being projected from the auxiliary to the physical
space. Denoting by $\langle {\bf p} | \phi({\bf k}) \rangle$ the ${\bf
p}$-Fourier transforms of specified electron wave functions, $A_e$ can
be expressed as: 
\begin{eqnarray}
A_e&({\bf p},& \omega) = \sum_{{\bf k}\sigma} \Big\{\sum_{\iota}
A^q_{\iota}({\bf k}, \omega) \Big[ \sum_{\nu} |\langle {\bf p} |
\phi^c_{\nu}({\bf k}) \rangle|^2 \nonumber \\ 
&\times& |U^{cq}_{\nu \iota}({\bf k})|^2 
+ {1\over N} \sum_{\lambda{\bf k}^{\prime} {\bf k}^{\prime\prime}}
\big[ |\langle {\bf p} | \phi^f_{\lambda}({\bf k}) \rangle|^2
U^{fq}_{\lambda\iota}({\bf k}^{\prime}, {\bf k})^* \nonumber \\ 
&\times& U^{fq}_{\lambda\iota}({\bf k}^{\prime\prime}, {\bf k}) \sqrt{
n^e_{\lambda}({\bf k}^{\prime}) + n^s_{\lambda,-\sigma}({\bf k}^{\prime}
- {\bf k})} \nonumber \\ 
&\times& \sqrt{ n^e_{\lambda}({\bf k}^{\prime\prime}) +
n^s_{\lambda,-\sigma}({\bf k}^{\prime\prime} - {\bf k})} + |\langle {\bf
p} |\phi^g_{\lambda}({\bf k}) \rangle|^2 \nonumber \\ 
&\times& U^{gq}_{\lambda\iota}({\bf k}^{\prime}, {\bf k})^*
U^{gq}_{\lambda\iota}({\bf k}^{\prime\prime}, {\bf k}) \sqrt{
n^h_{\lambda}({\bf k}^{\prime}) + n^s_{\lambda\sigma}({\bf k} - {\bf
k}^{\prime})} \nonumber \\ 
&\times& \sqrt{ n^h_{\lambda}({\bf k}^{\prime\prime}) +
n^s_{\lambda\sigma}({\bf k} - {\bf k}^{\prime\prime})} \big] \Big]
\nonumber \\ 
&+& {1 \over N} \sum_{\lambda\mu {\bf k}^{\prime}} |\langle {\bf p} |
\phi^{\zeta p}_{\lambda\mu}({\bf k}) \rangle|^2 \int d\omega^{\prime}
A^p_{\mu}({\bf k}^{\prime}, \omega^{\prime}) \nonumber \\ 
&\times& [-\cosh{^2(\xi_{\lambda,{\bf k} - {\bf k}^{\prime}})}
A^{\zeta}_{\lambda}({\bf k} - {\bf k}^{\prime}, \omega -
\omega^{\prime})\nonumber \\ 
&+& \sinh{^2(\xi_{\lambda,{\bf k} - {\bf k}^{\prime}})}
A^{\zeta}_{\lambda}({\bf k} - {\bf k}^{\prime}, \omega^{\prime} -
\omega)] \nonumber \\ 
&\times& [f_{_T}(\omega^{\prime}) + b_{_T}(\omega^{\prime} - \omega)]
\Big\}. 
\end{eqnarray} 
Note that $A_e$ is expressed here in terms of $A^q_{\iota}$ and
$A^{\zeta}_{\lambda}A^p_{\mu}$, corresponding to diagonal elements of
the auxiliary space Green's functions in the basis of the eigenstates of
the bare auxiliary fields. The electron bands are based on eigenstates
of the coupled-fields system, where ${\cal H}^{\prime}$ (4) introduces
hybridization between QE ($\iota$) and svivon-stripon ($\lambda\mu$)
states. 

The $n^e_{\lambda} + n^s_{\lambda\sigma}$ ($n^h_{\lambda} +
n^s_{\lambda\sigma}$) terms in Eq.~(21) represent a large-$U$ effect
discussed in Ref.~\cite{Eskes}. When all the sites are unoccupied by
large-$U$ electrons (holes), then $n^s_{\lambda\sigma}=0$,
$n^e_{\lambda}=1$ ($n^h_{\lambda}=1$), and the above terms contribute a
factor one in $A_e$, reflecting the fact that there is spectral weight
for both spin states per site in the lower (upper) Hubbard band. On the
other hand, when each site is occupied by a large-$U$ electron (hole),
then $n^s_{\lambda\sigma}=\half$ (for both spins), $n^e_{\lambda} =
n^h_{\lambda}=0$, and the above terms contribute factors half in $A_e$,
reflecting the fact that there is spectral weight for one electron
(hole) state per site, in both the lower and the upper Hubbard bands. 

An important question is how the quasi-continuum of QE bands is
projected into physical electron bands. The eigenvector elements
$U^{fq}_{\lambda\iota}$ and $U^{gq}_{\lambda\iota}$ appearing in
Eq.~(21) are of order $1/\sqrt{N}$, and their phases are quite random.
Consequently almost all the QE ($\iota$) bands have a contribution of
order $1/N$ to an ``incoherent'' background of $A_e$. A contribution of
``coherent'' bands to $A_e$, in the vicinity of $E_{_{\rm F}}$ comes
from few QE bands for which $U^{cq}_{\nu \iota}\sim 1$, or for which
almost all $U^{fq}_{\lambda\iota}$ or $U^{gq}_{\lambda\iota}$ have the
same phase. These bands correspond to QE's which are closely related to
physical electrons. 

Eq.~(21) preserves the auxiliary-space $\omega$ dependencies (10--20)
for the physical spectral functions (both in the coherent bands, and in
the incoherent background). The results for $\Gamma^q$ in Eq.~(13) are
reflected in non-Fermi-liquid linewidths, having a $\propto\omega$ and a
constant term, as detected in ARPES results \cite{Valla}. The convoluted
stripon--svivon term in Eq.~(21) hybridizes both with the incoherent
background, and the coherent bands. Since this term (in $A_e$) is
similar to the expression for $\Gamma^q$ in Eq.~(5), its $\omega$
dependence has [similarly to Eq.~(13)] a $\propto\omega$ and a constant
term, as that of the QE contribution to $A_e$ [given by Eq.~(13)]. 

The fact that the stripons introduce lower periodicity to the QE's and
svivons, they are coupled to, is experimentally reflected in ``shadow
bands'' and other weak-intensity superstructure effects. The strongest
coupling is expected for svivons around ${\bf k}_0$ (see the behavior of
$\bar\epsilon^{\zeta}$ there in Fig.~3), and for QE's at BZ areas of
dominantly high electrons density of states (DOS), found in the LSCO,
BSCCO and TBCO systems around the ``antinodal'' points $({\pi \over a} ,
0)$ and $(0 , {\pi \over a})$. If (from its four possibilities) ${\bf
k}_0$ were chosen at $({\pi \over 2a} , {\pi \over 2a})$, then the areas
in the lattice BZ which the stripon states reside in (at least in the
above systems) would be around $\pm{\bf k}^p = \pm({\pi \over 2a} ,
-{\pi \over 2a})$ \cite{Comm3} (the antinodal points are at ${\bf k}^p
\pm {\bf k}_0$). Thus one finds in BZ ranges around the antinodal points
a stripon--svivon contribution to $A_e$ which is very close to $E_{_{\rm
F}}$ (at energies $\bar\epsilon^p \pm \bar\epsilon^{\zeta}$ of
considerable linewidths). Such behavior in antinodal areas has been
widely observed (see {\it e.g.} Ref.~\cite{Yoshida1}). The opening of an
SC gap decreases the linewidth of $\bar\epsilon^{\zeta}$ near ${\bf
k}_0$ (see Fig.~3) resulting in narrow antinodal $\bar\epsilon^p \pm
\bar\epsilon^{\zeta}$ states, as will be discussed below. 

The renormalization of the QE energies, shown in Fig.~2, is expected to
occur for electron bands projected from them. Thus the experimentally
observed band slopes are smaller than the LDA predictions. Also, the
effect of the logarithmic singularity in Fig.~2 has been observed in
ARPES as a ``kink'' \cite{Lanzara,Johnson} in the ``nodal'' band as
$E_{_{\rm F}}$ is approached from below [near point $({\pi \over 2a} ,
{\pi \over 2a})$ in the BZ]. But it was attributed to coupling to
phonons \cite{Lanzara} or to the neutron scattering resonance mode
\cite{Johnson}. Note, however, that such a coupling would generally
result in two opposite changes in the band slope (below and above the
coupled excitation energy) below $E_{_{\rm F}}$, while the experimental
kink looks more consistent with one change in slope below $E_{_{\rm
F}}$, as in Fig.~2. Also, QE's close to the nodal points are expected to
be coupled to svivons which are mostly fairly away from point ${\bf
k}_0$, and thus the inequalities (19--20), which are necessary for this
logarithmic singularity to occur, are significant. 

This kink was not found in measurements in the n-type cuprate NCCO
\cite{Armitage}, which is consistent with the prediction here (suggested
by the author earlier \cite{Ashk1}) that in ``real'' n-type cuprates
this kink should be above, and not below $E_{_{\rm F}}$ (where ARPES
measurements are relevant). Also, there appears to be a sharp upturn in
the ARPES band in NCCO \cite{Armitage} very close to $E_{_{\rm F}}$
(believed there to be an artifact), which is expected here as the kink 
is approached from the other side of $E_{_{\rm F}}$ (see Fig.~2).

\section{THE NEUTRON RESONANCE MODE}

Spectroscopies measuring spin-flip excitations are largely determined by
terms like $\langle s_{i\sigma}^{\dagger} s_{i,-\sigma}
s_{j,-\sigma}^{\dagger} s_{j\sigma} \rangle$, contributing the
following term to the imaginary part of the spin susceptibility at wave
vector ${\bf q}$ (within the lattice BZ) and energy $\omega$: 
\begin{eqnarray}
\chi^{\prime\prime}({\bf q}, \omega) &\sim& \sum_{{\bf k}}
\sinh{(2\xi_{{\bf k}})} \sinh{(2\xi_{{\bf q} - {\bf k}})} \nonumber \\ 
&\times& \int d\omega^{\prime} A^{\zeta}({\bf k}, \omega^{\prime})
\big\{ A^{\zeta}({\bf q} - {\bf k}, -\omega -\omega^{\prime}) \nonumber
\\ 
&-& A^{\zeta}({\bf q} - {\bf k}, \omega - \omega^{\prime}) +
2A^{\zeta}({\bf q} - {\bf k}, \omega^{\prime} - \omega) \nonumber \\ 
&\times& [b_{_T}(\omega^{\prime} - \omega) - b_{_T}(\omega^{\prime}]
\big\}. 
\end{eqnarray} 
The effect of the negative minimum of $\bar\epsilon^{\zeta}({\bf k})$ at
${\bf k}_0$, in Fig.~3, and especially in the SC state, where its
linewidth is small, is the existence of a peak in
$\chi^{\prime\prime}({\bf q}, \omega)$ at ${\bf q} = 2{\bf k}_0 = {\bf
Q}$ (the AF wave vector) and $\omega = -2\bar\epsilon^{\zeta}({\bf
k}_0)$. 

This peak is consistent with the neutron-scattering resonance [thus
$E_{\rm res} = -2\bar\epsilon^{\zeta}({\bf k}_0)$] found in the
high-$T_c$ cuprate at $\sim 0.04\;$eV \cite{Rossat,Fong,Bourges1,Dai}.
The observation that the energy of the neutron resonance mode has a
local maximum at ${\bf k}={\bf Q}$ is also consistent with Fig.~3 and
Eq.~(22). However, also a peak whose energy is rising with ${\bf k} -
{\bf Q}$ is expected due to the range where $\bar\epsilon^{\zeta}({\bf
k})$ is positive and rising. And indeed, recent measurements
\cite{Reznik1} show neutron-scattering peak branches dispersing both
downward and upward with approximate circular symmetry around ${\bf
k}={\bf Q}$, as expected here. The incommensurate low-energy
neutron-scattering peaks, corresponding to the stripe-like
inhomogeneities \cite{Tran}, occur at points ${\bf Q} \pm 2{\bf q}$, in
directions where $\bar\epsilon^{\zeta}({\bf k}_0 \pm {\bf q}) = 0$, and
the slope of $\bar\epsilon^{\zeta}({\bf k})$ at ${\bf k}_0 \pm {\bf q}$
(see Fig.~3) is not too steep. 

Except for the single-layer TBCO, the resonance mode has been observed
in bilayer cuprates, where its symmetry is odd with respect to the
layers exchange. For two exchange-coupled CuO$_2$ planes \cite{Comm1},
the bare spinon energies split into ``acoustic'' (with a V-shape zero
minimum) and ``optical'' bands, of odd and even symmetries,
respectively. In an AF phase, double-spinon excitations of these modes
result in acoustic and optical spin-wave modes, as has been observed
\cite{Reznik2}. The self energy renormalization, resulting in a negative
minimal svivon energy, demonstrated in Fig.~3, is relevant to the (odd)
acoustic mode [where $\bar\epsilon_o^{\zeta}({\bf k}_0)<0$], while the
renormalization of the (even) optical mode is a minor shift, resulting
in a minimal energy $\bar\epsilon_e^{\zeta}({\bf k}_0)>0$. Generalizing
Eq.~(22) to the bilayer case yields the neutron resonance peak for odd
symmetry, around energy $E_{\rm res} = -2\bar\epsilon_o^{\zeta}({\bf
k}_0)$ at ${\bf k} = {\bf Q}$, as discussed above, while for even
symmetry one expects a higher energy peak, whose minimal energy
$2\bar\epsilon_e^{\zeta}({\bf k}_0)$ is at ${\bf k} = {\bf Q}$. This is
consistent with high-energy neutron-scattering measurements
\cite{Bourges2}. 
    
Thus, the neutron resonance mode is an excitation towards the
destruction of the stripe-like inhomogeneities, and its width determines
the speed of their dynamics. $E_{\rm res}/2$ is an indicator of the
low-energy scale. Transport properties, and especially thermoelectric
power, are sensitive to another indicator of the low-energy scale, the
stripon bandwidth. Expressions for them were worked out in
Ref.~\cite{Ashk1}, from which also this bandwidth was found to be
$\omega^p \sim 0.02\;$eV. 

\section{STRIPON HOPPING PROCESSES}

A major consequence of the coupling between the stripon, svivon, and QE
fields (worked out above) is stripon hopping between different charged
stripes, through intermediary QE+svivon states. Adiabatic snapshots of
consecutive steps in such inter-stripe hopping scenarios, within the
same section of stripe-like inhomogeneity as in Fig.~1, are shown in
Figs.~4,5. Dynamics of carriers beside those demonstrating the hopping
steps is ignored for simplicity. 

\begin{figure}[b!] 
\centerline{\epsfig{file=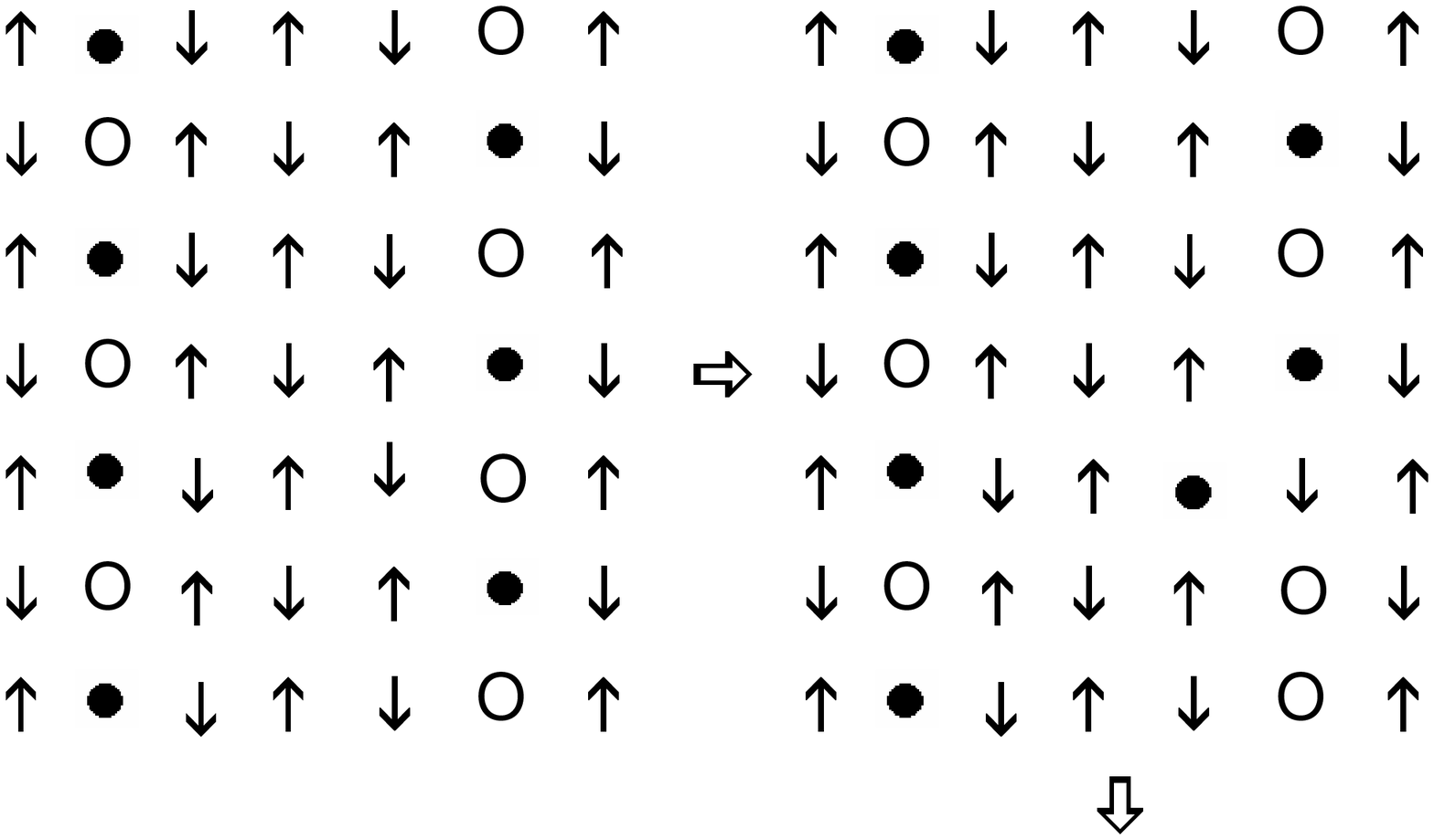,width=3.25in}}
\centerline{\epsfig{file=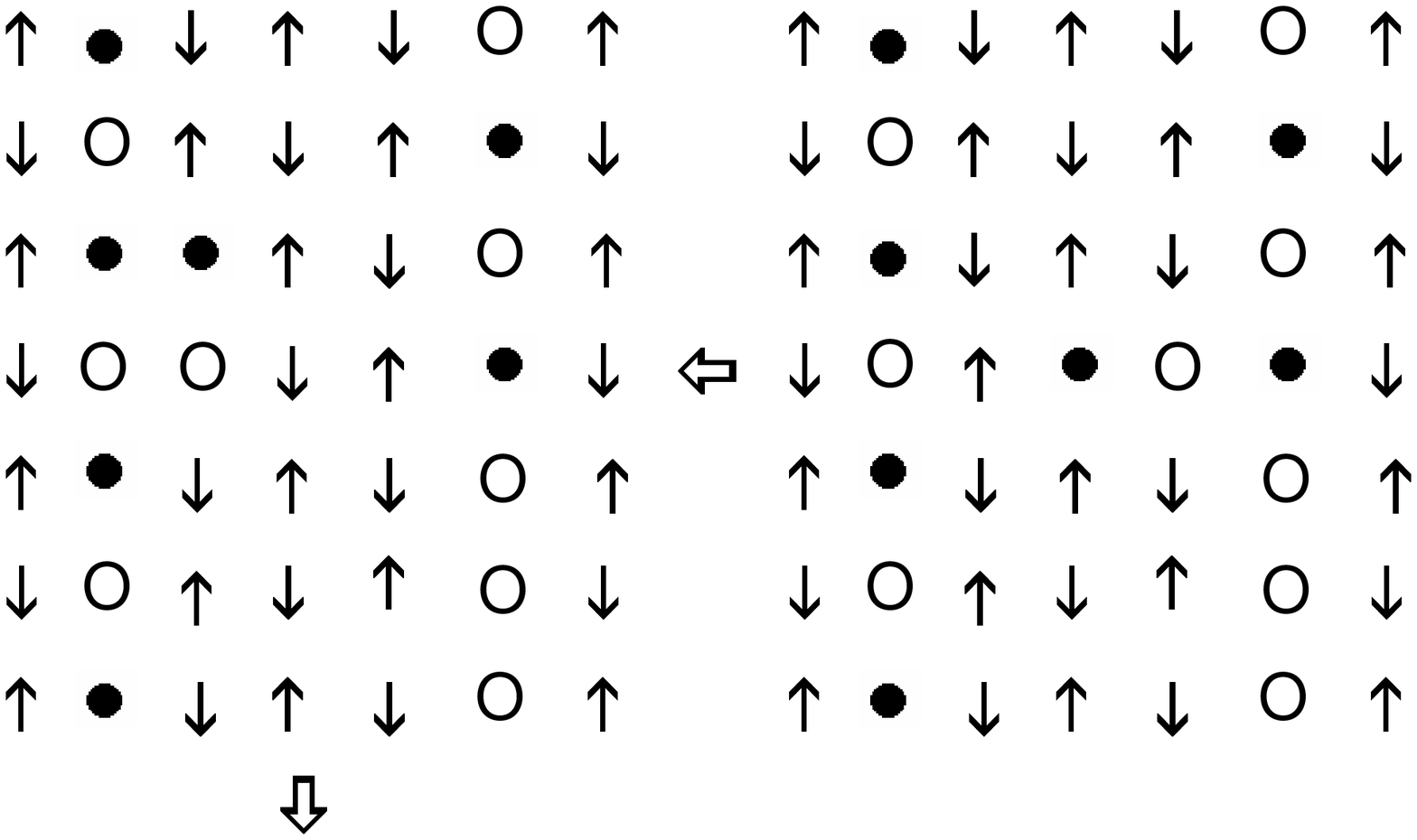,width=3.25in}}
\centerline{\epsfig{file=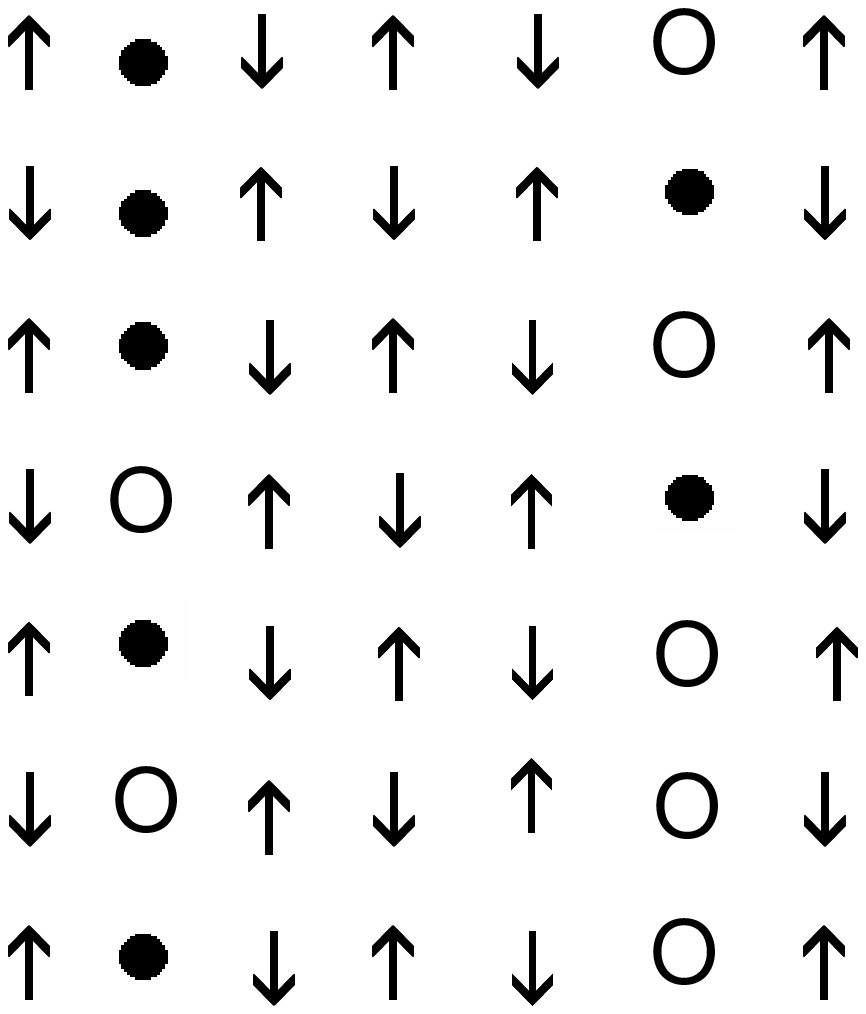,width=3.25in}}
\vspace{10pt}
\caption{Adiabatic snapshots of consecutive steps in a scenario for
``$t^{\prime}$-induced'' inter-stripe stripon hopping via QE+svivon
states.} 
\label{fig4}
\end{figure}

Within the $t$--$t^{\prime}$--$J$ model inter-site hopping could be
either between nearest neighbors ($t$-induced), or between next nearest
neighbors ($t^{\prime}$-induced). The inter-stripe stripon hopping
scenarios illustrated in Fig.~4 and Fig.~5 are through four consecutive
$t^{\prime}$-induced and $t$-induced inter-site hopping steps,
respectively. The intermediary states are of QE's and svivons, where the
latter propagate through spin (and lattice) dynamics. 

There is a difference between the dependencies of these two types of
inter-stripe hopping processes on the stripe structure. The
$t^{\prime}$-induced hopping could occur in a similar manner to that
illustrated in Fig.~4 also if the AF stripes were wider, and also in the
case of ``diagonal stripes'', existing in lightly doped (non-SC)
cuprates. On the other hand, $t$-induced hopping, as illustrated in
Fig.~5, is specific to the shown case of four-sites separation between
the charged stripes, and would involve higher energy intermediary states
for wider AF stripes, or diagonal stripes. Thus $t$-induced inter-stripe
hopping should be less effective for low doping levels. 

\begin{figure}[b!] 
\centerline{\epsfig{file=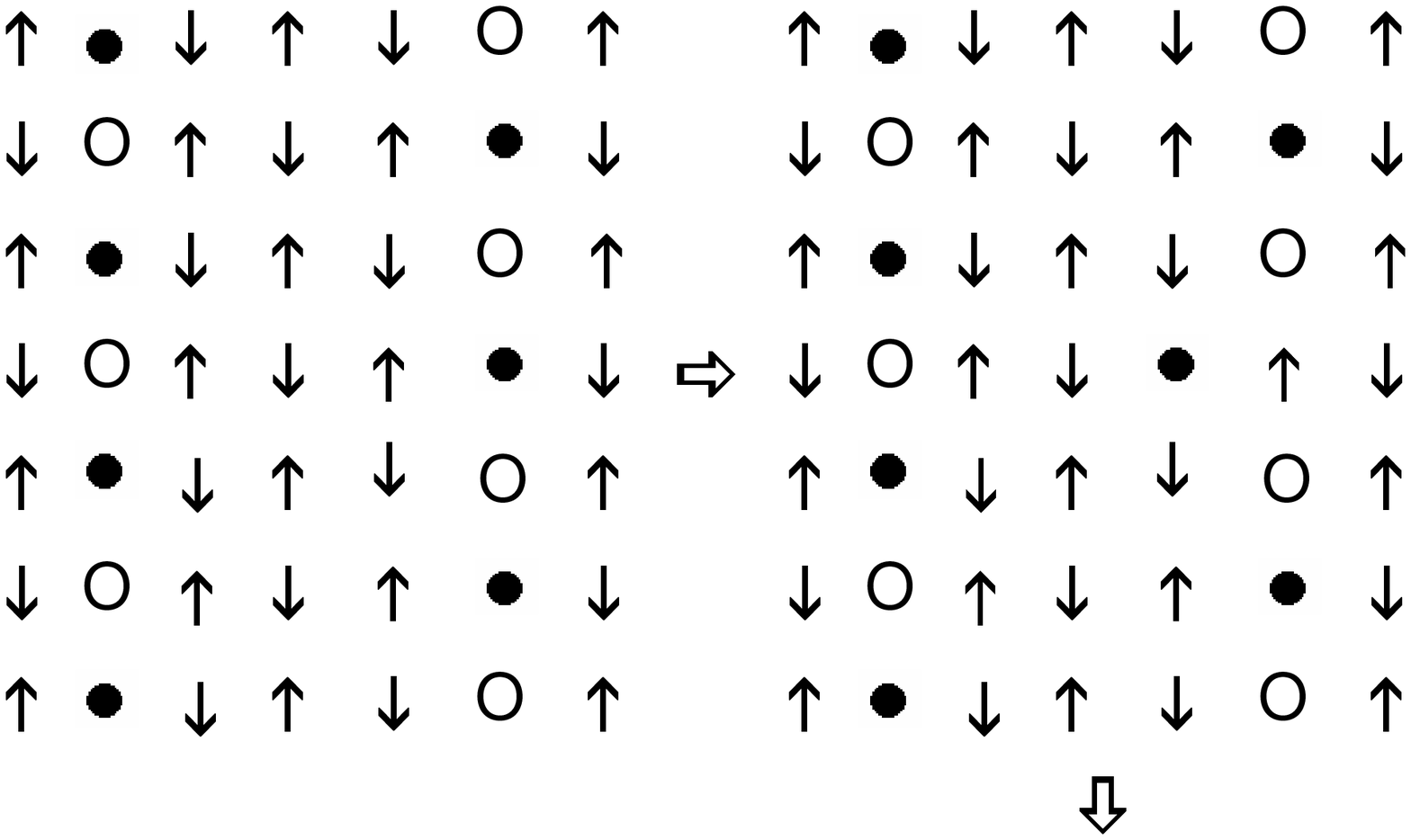,width=3.25in}}
\centerline{\epsfig{file=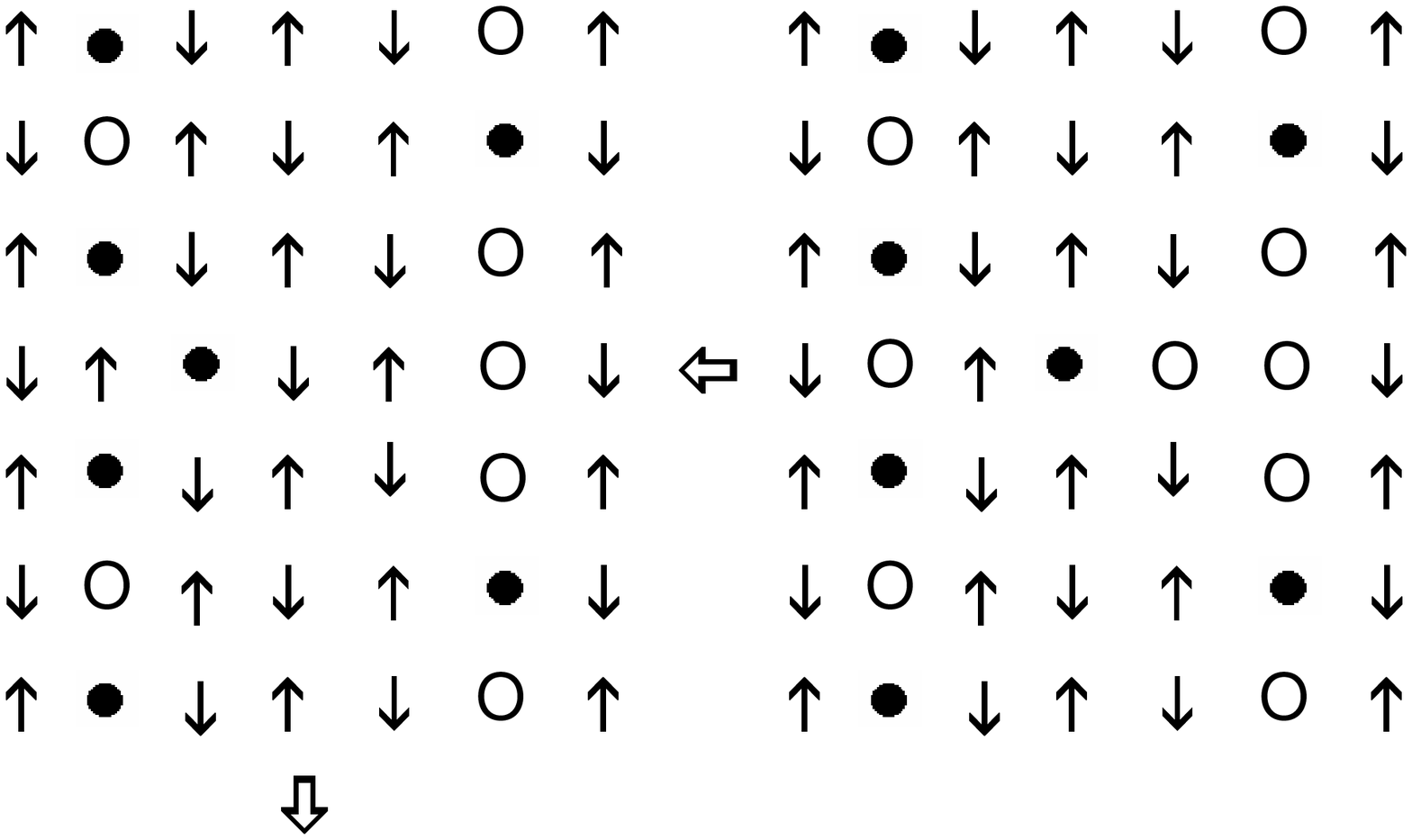,width=3.25in}}
\centerline{\epsfig{file=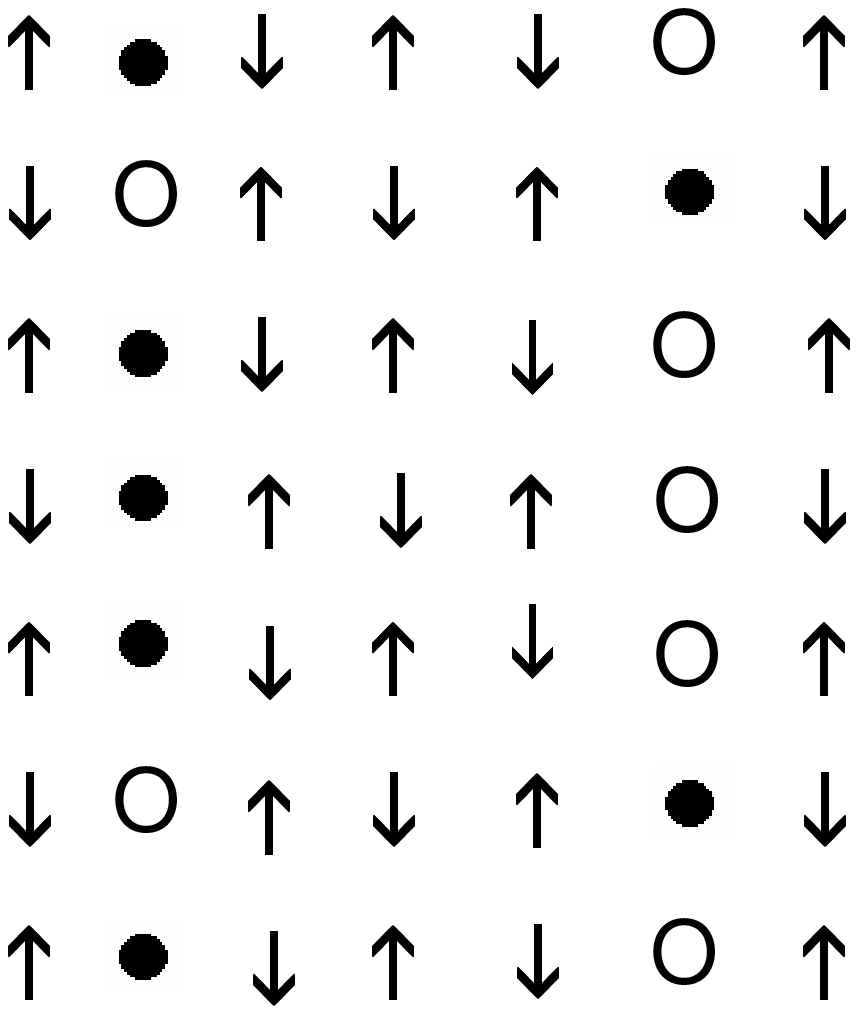,width=3.25in}}
\vspace{10pt}
\caption{Adiabatic snapshots of consecutive steps in a scenario for
``$t$-induced'' inter-stripe stripon hopping via QE+svivon states.} 
\label{fig5}
\end{figure}

This is consistent with the evolution of the metallic behavior of LSCO
with doping observed in ARPES measurements \cite{Yoshida2}. For low
doping levels $x$ the stripes are diagonal, and inter-stripe hopping is
mainly $t^{\prime}$-induced, resulting in the observed \cite{Yoshida2}
appearance of nodal states on $E_{_{\rm F}}$. For higher $x$ the stripes
become oriented as in Fig.~1, with the distance between charged stripes
being small enough for $t$-induced inter-stripe hopping to contribute
substantially. This results in the observed \cite{Yoshida2} appearance
of antinodal states on $E_{_{\rm F}}$, and SC. 

\begin{figure}[b!] 
\centerline{\epsfig{file=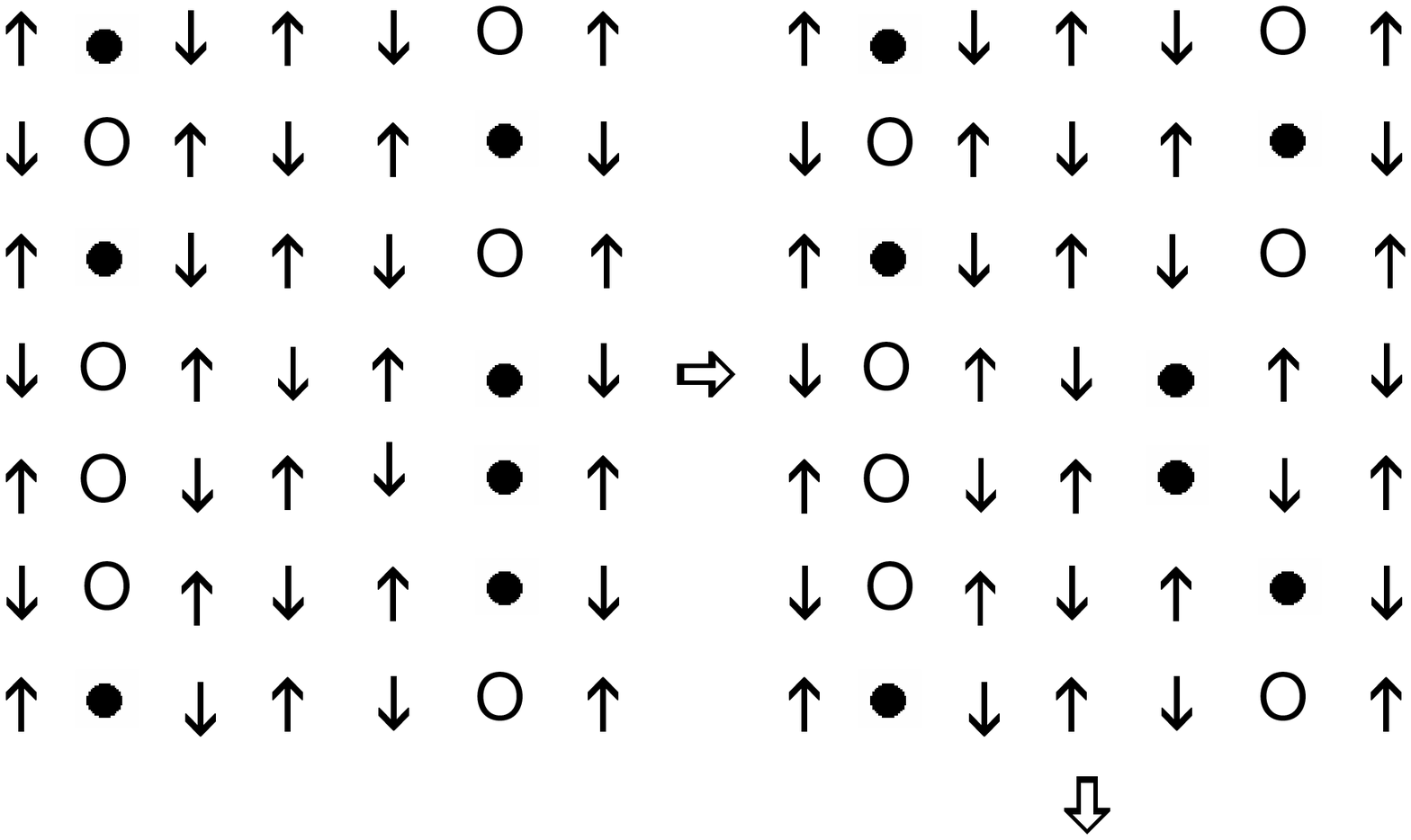,width=3.25in}}
\centerline{\epsfig{file=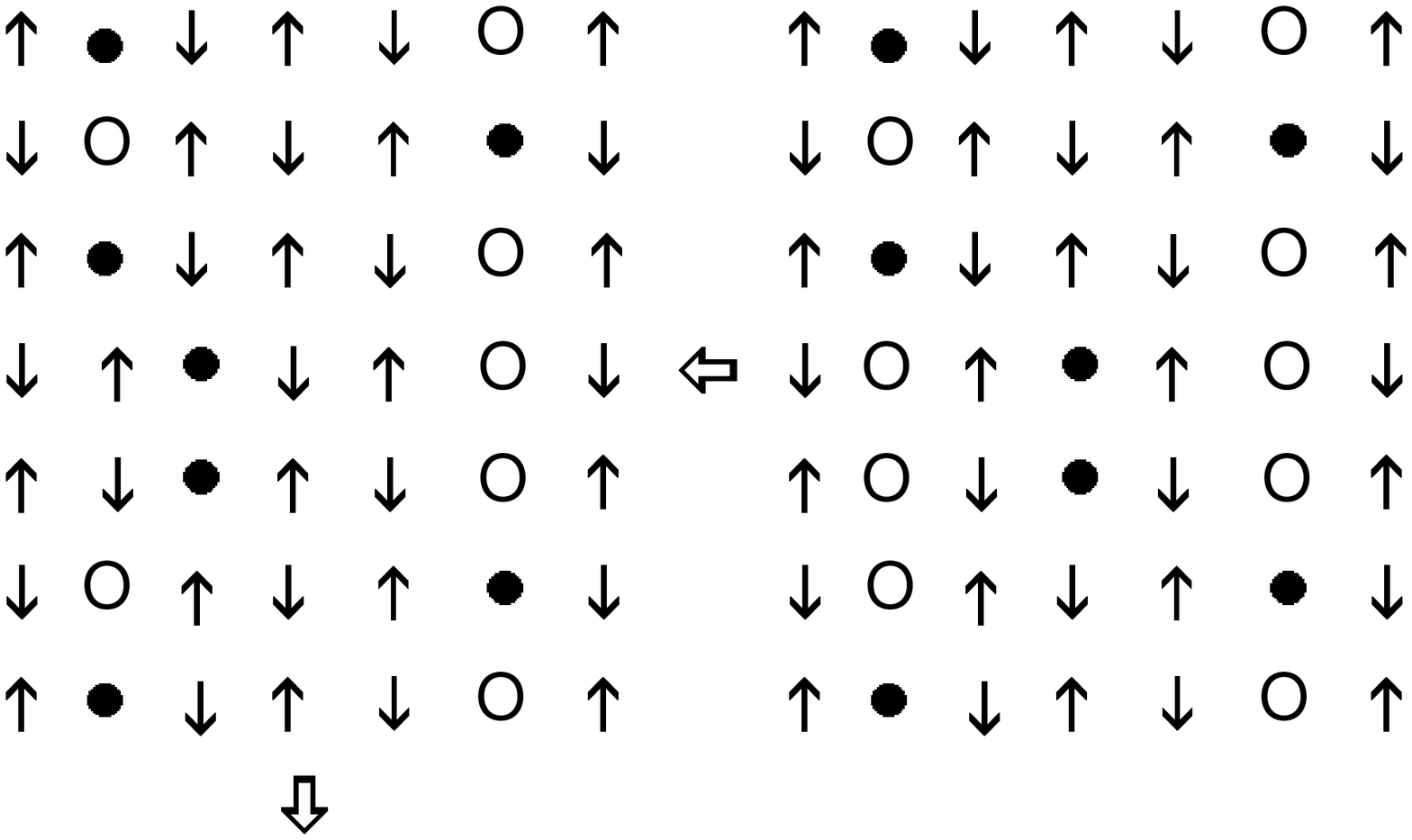,width=3.25in}}
\centerline{\epsfig{file=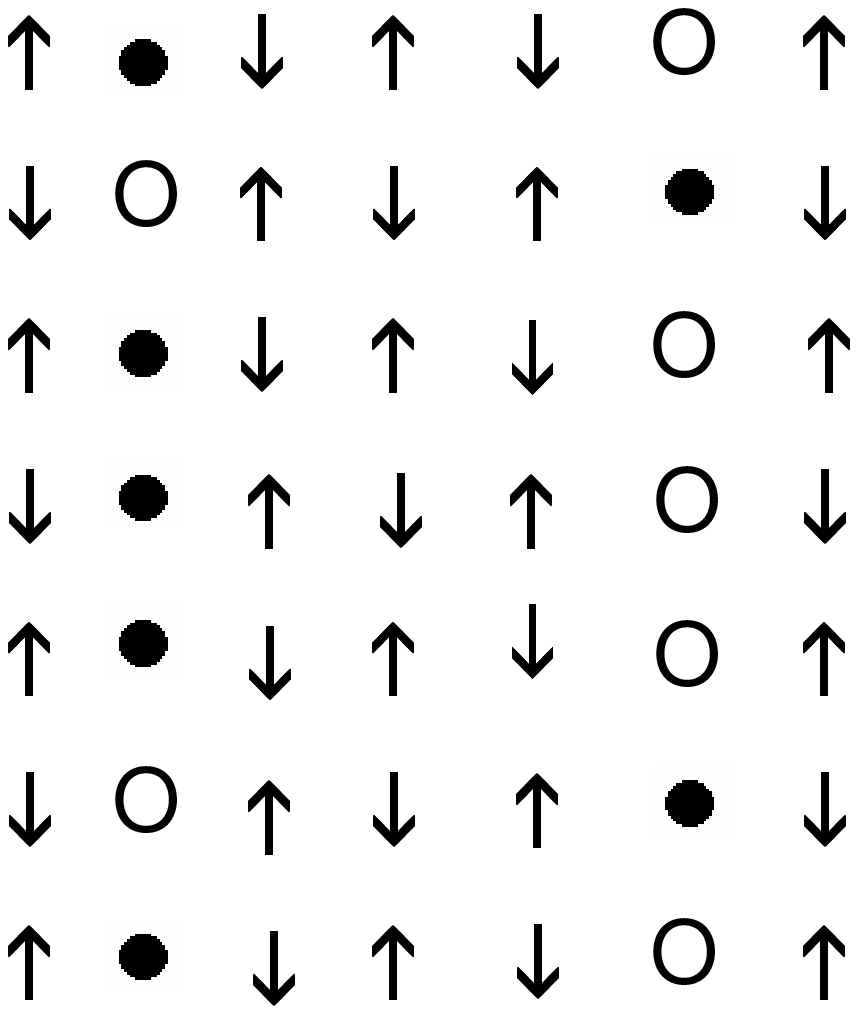,width=3.25in}}
\vspace{10pt}
\caption{Adiabatic snapshots of consecutive steps in a scenario for
inter-stripe hopping of stripon pairs via QE pair states.} 
\label{fig6}
\end{figure}

\section{HOPPING-INDUCED PAIRING}

It has been pointed out by the author \cite{Ashk1,Ashk3} that the
electronic structure discussed here provides pairing due to transitions
between pair states of stripons and QE's through the exchange of svivons
(the pairing diagram is sketched in Ref.~ \cite{Ashk3}). A scenario for
inter-stripe pair hopping, contributing to such pairing, is illustrated
in Fig.~6. The starting (ending) steps in this pair-hopping scenario is
a $t$-hopping of a nearest-neighbor ``dressed'' stripon pair, followed
(preceded) by turning this pair into (from) a pair of nearest-neighbor
QE's, through the exchange of a svivon [consisting of turning a pair of
nearest-neighbor opposite spins into (from) a singlet pair of
fluctuating spins]. The two intermediary steps consist of hopping of the
nearest-neighbor QE pair, which could be either $t^{\prime}$-hopping or
$t$-hopping with nearest-neighbor spin-flip \cite{Comm4}. Note
that $t^{\prime}$-hopping allows also scenarios where the QE pair gets
apart and joins again with no svivon excitation. Also note that such
pairing scenarios do not occur for diagonal stripes. 

Such pair-hopping scenarios result in gain in inter-stripe hopping
energy, compared to the single-particle inter-stripe hopping scenarios,
sketched in Figs.~4,5, avoiding intermediary svivon excitations.
Furthermore, hybridization with other orbitals, beyond the
$t$--$t^{\prime}$--$J$ model, results in further gain in (both
intra-plane and inter-plane) hopping energy, since it enables the QE
pair to move from the CuO$_2$ plane section shown in Fig.~6, to another
section which may be in another CuO$_2$ plane. 

The pairing diagram \cite{Ashk3} leads to Eliasherg-type equations, of
coupled stripon and QE pairing order parameters. Coherent pairing should
be between two subsets of the QE's and stripons. For QE's these subsets
are, naturally, chosen to be those of the spin-up ($\uparrow$) and
spin-down ($\downarrow$) QE's. As was illustrated in Fig.~1, for a
CuO$_2$ plane within the $t$--$t^{\prime}$--$J$ model the $\uparrow$
bare QE's can reside on $\downarrow$ sites, and the $\downarrow$ bare
QE's can reside on $\uparrow$ sites of the stripe-like inhomogeneities. 

An adiabatic snapshot of an extended section of a stripe-like
inhomogeneity, including an expected crossover between stripe segments
directed in the $a$ and the $b$ directions, is shown in Fig.~7. Denoted
are the available sites for the $\uparrow$ and $\downarrow$ QE subsets.
The stripons are spinless, but since the QE subsets have a spatial
interpretation in the CuO$_2$ planes, within the adiabatic time scale,
it is logical to choose the stripon subsets also on a spatial basis, in
a manner which optimizes pairing through scenarios similar to that
illustrated in Fig.~6. Thus the pairing subsets are chosen for bare
stripons such that the nearest neighbors (on a charged stripe) of a site
corresponding to one subset are sites corresponding to the other subset.
These subsets are denoted by $\bigtriangleup$ and $\bigtriangledown$,
and the sites available for them are shown in Fig.~7 too. 

The pairing order parameters are the QE and stripon pair-correlation
functions defined [in the position (${\bf r}$) representation] as: 
\begin{eqnarray}
\Phi^q({\bf r}_1,{\bf r}_2)  &\equiv& \langle q^{\dagg}_{\uparrow}({\bf
r}_1) q^{\dagg}_{\downarrow}({\bf r}_2) \rangle, \\ 
\Phi^p({\bf r}_1,{\bf r}_2)  &\equiv& \langle p^{\dagg}_{\bigtriangleup}
({\bf r}_1) p^{\dagg}_{\bigtriangledown}({\bf r}_2) \rangle. 
\end{eqnarray}
The coupled Eliashberg-type equations for these order parameters can be
expressed (omitting band indices for simplicity), in the position and
the Matzubara ($\omega_n$) representations, as: 
\begin{eqnarray}
\Phi^q&({\bf r}_1&,{\bf r}_2,i\omega_n) = \sum_{n^{\prime}} \int d{\bf
r}_1^{\prime} \int d{\bf r}_2^{\prime} \nonumber \\ 
&\times& K^{qp}({\bf r}_1,{\bf r}_2,n; {\bf r}_1^{\prime},{\bf
r}_2^{\prime},n^{\prime}) \Phi^p({\bf r}_1^{\prime},{\bf
r}_2^{\prime},i\omega_n^{\prime}), \\ 
\Phi^p&({\bf r}_1&,{\bf r}_2,i\omega_n) = \sum_{n^{\prime}} \int d{\bf
r}_1^{\prime} \int d{\bf r}_2^{\prime} \nonumber \\ 
&\times& K^{pq}({\bf r}_1,{\bf r}_2,n; {\bf r}_1^{\prime},{\bf
r}_2^{\prime},n^{\prime}) \Phi^q({\bf r}_1^{\prime},{\bf
r}_2^{\prime},i\omega_n^{\prime}).
\end{eqnarray} 
Expressions for the kernel functions $K^{qp}$ and $K^{pq}$ are obtained
from the pairing diagrams. They depend on $\Phi^q$ and $\Phi^p$ up to
the temperature where the latter vanish \cite{Comm5}. 

\begin{figure}[b!] 
\centerline{\epsfig{file=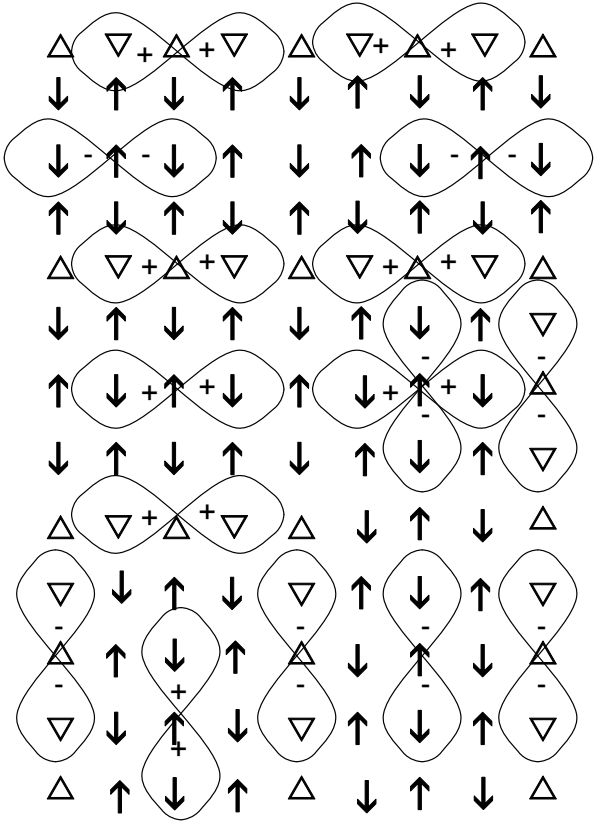,width=3.25in}}
\vspace{10pt}
\caption{An adiabatic snapshot of an extended section of a stripe-like
inhomogeneity, where the available sites for the QE and stripon pairing
subsets are illustrated, as well as sketches demonstrating the local
symmetry of the pairing order parameters $\Phi^q$ and $\Phi^p$.} 
\label{fig7}
\end{figure}

Sketches demonstrating the local symmetry (within the adiabatic time
scale) of $\Phi^q({\bf r}_1,{\bf r}_2)$ and $\Phi^p({\bf r}_1,{\bf
r}_2)$, on the basis of Eqs.~(23--26), are illustrated in Fig.~7. In
these sketches point ${\bf r}_1$ is fixed on selected sites (of the
$\uparrow$ and $\bigtriangleup$ QE and stripon subsets), while point
${\bf r}_2$ is varied over the space including the nearest neighbors.
Two types of local sign reversals of the order parameters are observed.
One is for $\Phi^q$ on different sides of a charged stripe (serving as
an anti-phase domain wall also regarding pairing). If two QE sites on
different sides of a charged stripe have a stripon site midway between
them, then one of them is $\uparrow$ and the other is $\downarrow$, and
since the exchange of the two fermion operators in the definition of
$\Phi^q$ in Eq.~(23) results in sign reversal, there must be sign
reversal in $\Phi^q$ between the two sides. The other sign reversal is
(both for $\Phi^q$ and $\Phi^p$) between $a$-oriented and $b$-oriented
stripe segments meeting in a ``corner'' (shown in Fig.~7). This sign
reversal provides optimal pairing energy, especially in the corner
regions, yielding maximal $|\Phi^q({\bf r}_1,{\bf r}_2)|$ when ${\bf
r}_1$ and ${\bf r}_2$ are at nearest neighbor QE sites \cite{Comm6}
(where the QE's have opposite spins and can pair), and zero $\Phi^q({\bf
r}_1,{\bf r}_2)$ when ${\bf r}_1$ and ${\bf r}_2$ are at next nearest
neighbor QE sites (where the QE's have the same spin and cannot pair). 
 
The symmetry features of $\Phi^q$ and $\Phi^p$, discussed above, are
reflected in symmetry of the physical pairing order parameter. The
overall symmetry is of $d_{x^2-y^2}$ type. However the sign reversal of
$\Phi^q$ through the charged stripes, combined with the dynamics of the
stripe-like inhomogeneities, and the lack of coherence between their
details in different CuO$_2$ planes (at least in ones which are not
adjacent) would result in experimental observations which are not always
consistent with $d_{x^2-y^2}$-wave pairing. The experimental conclusions
concerning the gap symmetry are still somewhat controversial, though
there is strong support in the existence of features of
$d_{x^2-y^2}$-wave pairing. Note that the possibility of symmetry
mixtures associated with broken time-reversal symmetry (which seem to be
observed experimentally) is not addressed here. 

The QE and stripon pairing gaps $2\Delta^q$ and $2\Delta^p$ are closely
related to $\Phi^q$ and $\Phi^p$, and have the same symmetries. The
electronic pairing gap $2\Delta$ equals $2\Delta^q$; they vanish at the
nodal points and equal $2\Delta_{\rm max}$ at the antinodal points.
$\Delta^p$ is almost constant (it is reminded that the stripon states
reside around points $\pm{\bf k}^p$ of the lattice BZ, defined above).
$\Delta^p$ is greater than $\Delta^q_{\rm max}$, and they are both
expected to be higher when the AF/stripes effects are stronger, thus
decrease with the doping level $x$. However the value of $\Delta^p$ is
limited by the condition: 
\begin{equation}
\Delta^p - {E_{\rm res} \over 2} = \Delta^p + \bar\epsilon^{\zeta}({\bf
k}_0) \simeq \Delta^q_{\rm max} = \Delta_{\rm max}, 
\end{equation}
since above this value unpaired stripons would be formed from unpaired
QE's and svivons. Thus one expects a decrease of $\Delta_{\rm max}$ with
$x$, scaling with the pairing temperature ($T_{\rm pair}$) line in
Fig.~8 (approximately according to the BCS factor \cite{Comm5}) as has
been observed. 

On the other hand, $-\bar\epsilon^{\zeta}({\bf k}_0)$, which is zero for
an AF, has a tendency to increase distancing from an AF state, as $x$
rises. However, one must have: 
\begin{equation}
{E_{\rm res} \over 2} = - \bar\epsilon^{\zeta}({\bf k}_0) < \Delta_{\rm
max}, 
\end{equation}
otherwise svivons around ${\bf k}_0$ would have enough energy to break
pairs and, consequently, be scattered to QE-stripon states. Thus the
value of $-\bar\epsilon^{\zeta}({\bf k}_0)$ is expected to cross over
from an increase to a decrease with $x$ (like $\Delta_{\rm max}$) when
their values get close to each other. And indeed, the neutron resonance
energy $E_{\rm res} = -2\bar\epsilon^{\zeta}({\bf k}_0)$ has been found
\cite{Bourges1} to cross over from an increase to a decrease with $x$
when its value gets close to $2\Delta_{\rm max}$ (in a manner resembling
the $T_c$ line in Fig.~8). 

\begin{figure}[b!] 
\centerline{\epsfig{file=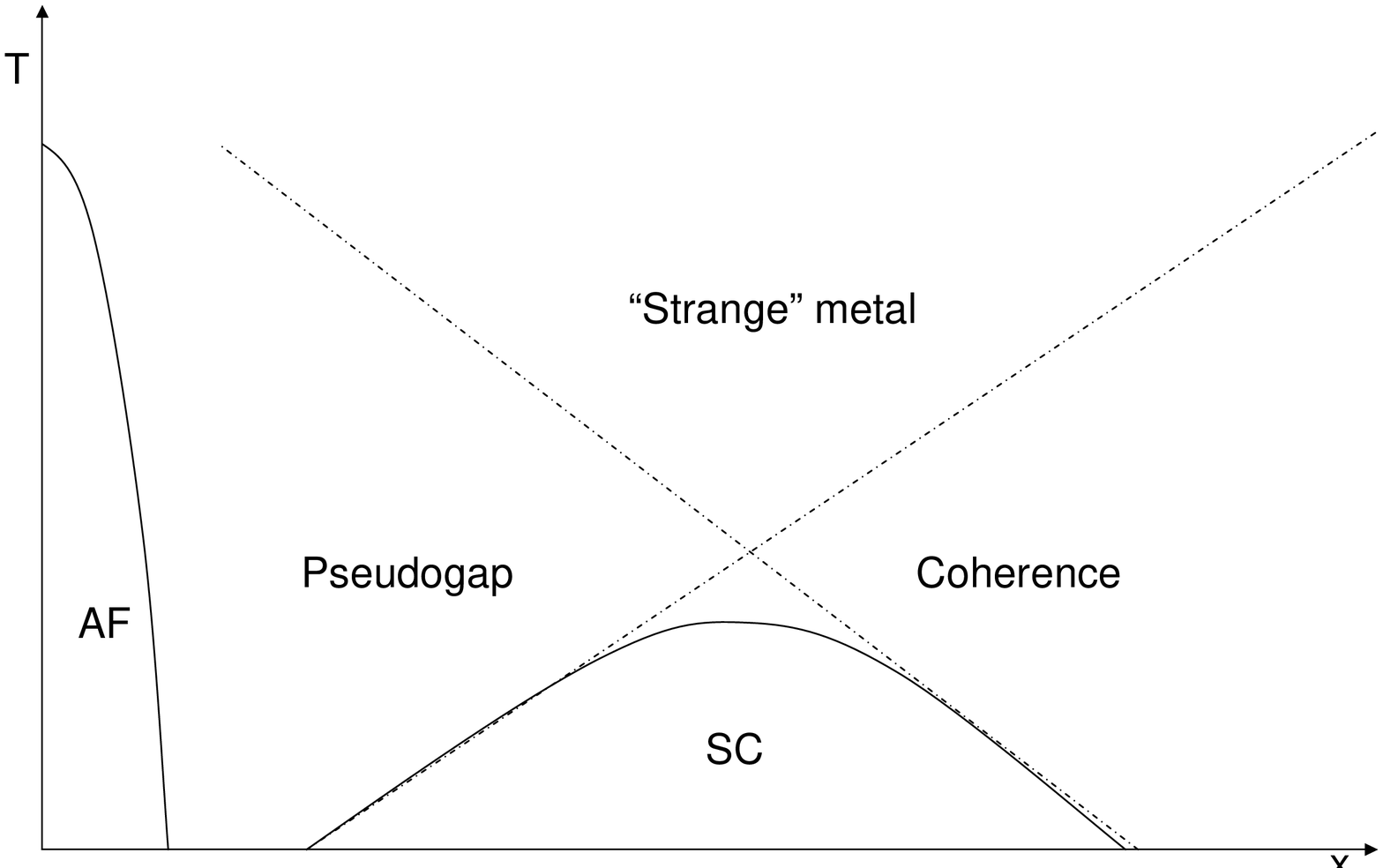,width=3.25in}}
\vspace{10pt}
\caption{A schematic phase diagram for the cuprates. The $T_c$ line is 
determined by the pairing line ($T_{\rm pair}$), decreasing with $x$,
and the coherence line ($T_{\rm coh}$), increasing with $x$ (broken
lines represent crossover regimes).} 
\label{fig8}
\end{figure}

\section{PAIRING AND COHERENCE}

The existence of SC requires not only the existence of pair
correlations, but also of phase coherence of the pairing order
parameters. Under conditions satisfied, for low $x$ values, within the
phase diagram of the cuprates, pairing occurs below $T_{\rm pair}$,
while SC occurs only below $T_{\rm coh} (< T_{\rm pair})$, where phase
coherence sets in. The normal-state pseudogap (PG), observed in the
cuprates above $T_c$ (except for high $x$ values) is a pair-breaking gap
at $T_{\rm coh} < T < T_{\rm pair}$ (see Fig.~8). Its size and symmetry
are similar to those of the SC gap, and specific heat measurements
\cite{Moca} imply that it accounts for most of the pairing energy. 

Pairing coherence requires energetical advantage of itineracy of the
electronic states near $E_{_{\rm F}}$ \cite{Comm2}. Thus $T_{\rm coh}$
is expected to increase with $x$ (as sketched in the coherence line in
Fig.~8), due to moving away from a Mott insulator, for which itineracy
is energetically suppressed. This coherence argument is independent of
pairing (which is energetically favorable without coherence in the PG
state) and continues to be valid also in the regime of the phase
diagram, shown in Fig.~8, where pairing does not exist (thus $T_{\rm
coh}> T_{\rm pair}$). Such a determination of $T_{\rm coh}$ is consistent
(in the regime where $T_{\rm coh} < T_{\rm pair}$) with a
phenomenological model \cite{Emery} evaluating $T_{\rm coh}$ on the
basis of the phase ``stiffness'' (maintaining its coherence against
fluctuations). It yields $T_{\rm coh} \propto n_s / m_s^*$, where
$m_s^*$ and $n_s$ are the effective pairs mass and density. This result
explains the ``Uemura plots'' \cite{Uemura} in the PG doping regime
(where $T_c = T_{\rm coh}$). 

Since the pair states here are fluctuating between QE and stripon pair
states, the pairs density $n_s$ is determined by the smaller of them,
thus by the density of stripon pairs. The thermoelectric power results
in Ref.~\cite{Ashk1} determine the occupancy of the stripon states to be
$50\%$ ($n^p=\half$) for slightly overdoped cuprates ($x\simeq 0.19$).
Thus the effective $n_s$ is maximal around this stoichiometry, being
determined by the density of hole-like stripon pairs for smaller $x$,
and of particle-like stripon pairs for larger $x$. This explains the
``boomerang-type'' behavior \cite{Niedermayer} of the Uemura plots
around $x\simeq 0.19$, which is a crossover regime for the value of
$T_c$ between those of $T_{\rm coh}$ in the underdoped regime, and of
$T_{\rm pair}$ in the heavily overdoped regime (see Fig.~8). 

Measurements of changes in the optical conductivity in the cuprates, as
temperature is lowered through $T_c$, reveal unconventional features,
which are different along the $c$-axis and perpendicular to it. The
$c$-axis effects \cite{marel1} below $T_c$ are characterized, beside the
opening of the SC gap, by the increase of the spectral weight at the
mid-IR range (well above the gap). This effect has been observed both in
bilayer and single-layer cuprates, and proposed \cite{marel1} to be the
signature of a $c$-oriented collective mode emerging (or sharpening)
below $T_c$. Within the theory presented here of transitions between
pair states of stripons and QE's, such a mode is consistent with a
$c$-axis plasmon-type mode of such coherent pairs (below $T_c$). These
pairs can hop in the $c$-direction, during their QE-pair stages, while
above $T_c$ $c$-axis hopping of stripons (through intermediary
QE--svivon states) is, at the most, limited to adjacent CuO$_2$ planes. 

Measurements of the in-plane optical conductivity \cite{marel2}, as
temperature is lowered through $T_c$, reveal, except for the overdoped
regime, the transfer of spectral weight from high energies (extending
over a broad range up to at least 2~eV), to the infrared range. This
behavior has been associated with the establishment of coherence
\cite{Norman}. Within the present approach \cite{Comm2} it is due to the
transfer of spectral weight of large-$U$ electrons from the high-energy
range, of the upper and lower Hubbard bands, to the intermediary and low
energy ranges, where they participate in the formation of itinerant
states based on QE's (where hybridization with small-$U$ electrons is
included), as was worked out in Eqs.~(10--18). 

This coherence effect is through the coherence line in the phase diagram
(see Fig.~8), and is not related to pairing. Let us regard its
manifestation through the shape of ARPES peaks close to $E_{_{\rm F}}$,
where pairing effects are not present. This is the case in nodal points
(where $\Delta^q = 0$), and the effect of coherence appears to be
\cite{Valla} the establishment of a sharp edge to the peaks. Furthermore,
in the overdoped regime coherence is expected to exist also (see
Fig.~8) in a temperature range above $T_c$ (and $T_{\rm pair}$). ARPES
measurements in this regime \cite{Yusof} show the appearance of
coherence effects at $T_{\rm pair}< T < T_{\rm coh}$, both in nodal and
antinodal peaks, consisting of the existence of a sharp peak edge and
resolved bilayer-split bands. 

\section{PAIR-BREAKING EXCITATIONS}

Both the effects of pairing and coherence are expected to exist below
$T_c$ in BZ ranges around the antinodal points. ARPES data from these
ranges \cite{Feng1,Feng2,Borisenko} as well as tunneling results
\cite{Renner,Kugler,Zasadzinski,Yurgens,Peter}, reveal a peak-dip-hump
structure, appearing below $T_c$, but missing considerably above it in
the PG regime. This structure is partly due to coherence-resolved
bilayer-split bands (discussed above), but the present approach explains
also another ``intrinsic'' aspect of this structure. 

Eq.~(21) distinguishes between QE and stripon-svivon contributions to
the spectral structure. The linewidth of the svivon band around ${\bf
k}_0$ (see Fig.~3) is expected to drop drastically below $T_c$, because
the appearance of a coherent pairing gap (and not a partial one as
in the PG regime) eliminates the scattering of these svivons to
QE-stripon states. This is observed in the width of the neutron
resonance peak at $E_{\rm res}= -2\bar\epsilon^{\zeta}({\bf k}_0)$,
becoming sharp only below $T_c$ (though its remanent may be observed in
the PG regime). Since the stripon bandwidth $\omega^p$ is expected
\cite{Ashk1} to be smaller than $\Delta^p$, the stripon-svivon
contribution (21) to the antinodal pair-breaking structure in the SC
state \cite{Comm3} is expected to consist, on each side of $E_{_{\rm
F}}$, of two peaks (and a dip between them) at distances $E_{\rm peak}
\gta \Delta^p \pm \bar\epsilon^{\zeta}({\bf k}_0)$ from $E_{_{\rm F}}$.
By Eq.~(27) the energies of these two (intrinsic) antinodal peaks can be
expressed as: 
\begin{equation}
E_{\rm peak,1} \gta \Delta_{\rm max}, \ \ \ \ \ 
E_{\rm peak,2} \gta \Delta_{\rm max} + E_{\rm res}.
\end{equation}

Beside these two intrinsic peaks there is (21) also a QE-derived
structure around the antinodal points. In bilayer cuprates it splits
(below $T_c$) into bonding (B) and anti-bonding (AB) bands. The B band
is generally not too close to the gap edge, and thus does not contribute
a sharp peak, but a wide hump. On the other hand, the AB band gets very
close to the gap edge (depending on $x$), and could, consequently, be
very sharp. At different stoichiometries it may be hard to distinguish
between the intrinsic peaks given in Eq.~(29) and the AB peak and B
hump, especially that they hybridize with each other. Tunneling results
\cite{Zasadzinski} on BSCCO, for different doping levels, support the
result (29) of two peaks separated by $E_{\rm res}$. The two peaks are
also observed in ARPES results on overdoped BSCCO \cite{Feng2}; however
the peak at $E_{\rm peak,1}$ has been identified there with that of the
AB band (which for the measured stoichiometry coincides with it, and
partly lies above $E_{_{\rm F}}$). This identification led to an
analysis \cite{Eschrig} under which there is only one intrinsic peak (at
$E_{\rm peak,2}$), resulting from coupling between the B and the AB
bands through the odd-symmetry neutron resonance mode. However, in
recent results \cite{Borisenko} on underdoped Pb-BSCCO, the peak at
$E_{\rm peak,1}$ is identified as an intrinsic one, while that of the AB
band (which for that stoichiometry lies fairly below $E_{_{\rm F}}$, and
is thus not very sharp) is identified as ``an AB hump''. 

The possibility to distinguish between the intrinsic peaks (29) and
QE-derived structure should be simpler in single-layer cuprates for
which less ``clean'' experimental results exist. Tunneling results in
single-layer BSCO \cite{Kugler} show a peak-dip-hump structure appearing
below $T_c$, and missing in the PG state. Recent tunneling results, for
different stoichiometries, in BSLCO show a hump in the PG state,
splitting below $T_c$ into two peaks, which are consistent with the
intrinsic peaks given in Eq.~(29). ARPES results above $T_c$ in BSLCO
\cite{Janowitz} reveal unexpected ``bilayer splitting'' in the antinodal
BZ area, and further measurements below $T_c$ would be helpful to
clarify whether this structure is due to the effect of the intrinsic
peaks. Recent ARPES measurements on LSCO thin films \cite{Pavuna} show
that $T_c$ rises under strain, while the bands close to $E_{_{\rm F}}$
become wider, and the topology of the Fermi surface around the antinodal
points is altered. Further higher resolution measurements of strain
effects on such films would be useful to study the antinodal excitations
structure. 

The spectral weight within the intrinsic peaks (29) depends on the
number of stripon-svivon states within them. The svivon states are those
around the minimum of $\bar\epsilon^{\zeta}$ at ${\bf k}_0$ (see
Fig.~3), and thus their number approximately scales with
$-\bar\epsilon^{\zeta}({\bf k}_0) = E_{\rm res}/2$, whose dependence on
$x$ \cite{Bourges1} approximately scales with the $T_c$ curve in Fig.~8.
The stripon states are within a narrow band, which splits in the SC
state \cite{Comm5}, through the Bogoliubov transformation, into two
halves (since $\omega^p < \Delta^p$), separated by $2\Delta^p$. Typical
states in these upper and lower bands are created, respectively, by
$p^{\dagger}_{\rm u}$ and $p^{\dagger}_{\rm l}$. They are expressed in
terms of creation and annihilation operators of stripons of the two
pairing subsets (24) through equations of the form ($|u|^2 + |v|^2 =
1$): 
\begin{equation}
p^{\dagger}_{\rm u} = u p^{\dagger}_{\bigtriangleup} + v
p^{\dagg}_{\bigtriangledown}, \ \ \ \ \ 
p^{\dagger}_{\rm l} = -v p^{\dagg}_{\bigtriangleup} + u
p^{\dagger}_{\bigtriangledown}.
\end{equation}
ARPES experiments at low temperatures measure the number of hole-like
stripon states in the lower band, which equals $\sum |v|^2$. By BCS
theory this number is $\propto (1-n^p)$, the occupancy of the stripon
states \cite{Ashk1} in the hole representation, which approximately
scales with $x$. 

Thus, the prediction for low-temperature ARPES measurements of the
$x$-dependence of spectral weight within the intrinsic peaks (29) is as
follows: an increase with $x$ is expected in the underdoped regime,
where the number of measured states of both svivons and stripons
increases with $x$; on the other hand, in the overdoped regime, where
the number of measured svivon states decreases with $x$, the increase
with $x$ of the measured spectral weight is expected to slow down, or
even turn into a decrease with $x$. Two reports have been published
\cite{Feng1} on such measurements in bilayer BSCCO, of the spectral
weight within the peak, omitting the background (including the
hump), and integrating over the antinodal BZ area. The resolution in
these measurements was not sufficient to distinguish between the two
peaks in Eq.~(29), or to distinguish them from the AB band peak. But
still they confirm the above prediction of an increase of the spectral
weight with $x$ in the underdoped regime, and its saturation, or
possible turning into a decrease with $x$ (both within the measurement
error bars) in the overdoped regime. This observation support the
possibility that the major contribution to the measured peak is the
intrinsic one [through Eq.~(29)]. 

\section{Conclusions}

High-$T_c$ superconductivity, and other puzzling physical properties of
the cuprates, are found to result from special conditions existing for
them in the regime of a Mott transition. Two features determine their
phase diagram: ({\it i}) hopping-induced pairing, which depends on
partial spin-charge separation, within dynamical stripe-like
inhomogeneities, becomes stronger at lower doping levels, closer to the
insulating side of the Mott transition; ({\it ii}) phase coherence,
which is necessary for superconductivity to occur, becomes stronger at
higher doping levels, closer to the metallic side of the Mott
transition. A non-Fermi-liquid approach is used, treating the
stripe-like inhomogeneities adiabaticlly. But it predicts their faster
dynamics in an un-paired state, thus restoring Fermi-liquid behavior in
the coherence regime.


\begin{thebibliography}{99}
\bibitem{Zaanen}J.~Zaanen, and O.~Gunnarsson, {\it Phys.~Rev.~B} {\bf
40}, 7391 (1989); K.~Machida, {\it Physica C} {\bf 158}, 192 (1989);
M.~Kato, {\it et al.}, {\it J.~Phys.~Soc.~Jpn} {\bf 59}, 1047 (1990);
V.~J.~Emery, and S.~A.~Kivelson, {\it Physica C} {\bf 209}, 597 (1993). 
\bibitem{Tran}J.~M.~Tranquada {\it et al.}, {\it Phys.~Rev.~B} {\bf 54},
7489, (1996); {\it Phys.~Rev.~Lett.} {\bf 78}, 338 (1997).
\bibitem{Andersen}O.~K.~Andersen {\it et al.}, {\it
J.~Phys.~Chem.~Solids} {\bf 56}, 1573 (1995). 
\bibitem{Barnes} S.~E.~Barnes, {\it Adv.~Phys.} {\bf 30}, 801 (1980). 
\bibitem{Ashk1}J.~Ashkenazi, {\it J.~Phys.~Chem.~Solids}, {\bf 63},
2277-2284 (2002); cond-mat/0108383; in the High-$T_c$ Cuprates'', in
{\it New Trends in Superconductivity}, edited by J.~F.~Annett and
S.~Kruchinin (Kluwer Academics Publishers, 2002), p. 51;
cond-mat/0203170. 
\bibitem{Ashk2}J.~Ashkenazi, {\it J.~Supercond.} {\bf 7}, 719 (1994).
\bibitem{Comm1}The treatment of the spinons in Ref.~\cite{Ashk2} is
based on an uncoupled CuO$_2$ plane in its basic periodicity. The
existence of AF order or stripe-like inhomogeneities introduces a lower
periodicity, resulting in some mixing between spinon states of different
${\bf k}$ values in the lattice BZ. The existence of exchange coupling
between adjacent CuO$_2$ planes results in a number of bare bands (as
the number of coupled CuO$_2$ planes), the lowest of which has a V-shape
zero minimum. 
\bibitem{Comm2}The existence of QE bands, based (at least partially) on
large-$U$ states, in the vicinity of $E_{_{\rm F}}$ requires the
transfer of such states from the upper and lower Hubbard bands to this
vicinity. Such a transfer occurs in the regime of a Mott transition,
within which the cuprates are believed to be, and it is increasing with
the doping level $x$. 
\bibitem{Ashk3}J.~Ashkenazi, {\it High-Temperature Superconductivity},
edited by S.~E.~Barnes, J.~Ashkenazi, J.~L.~Cohn, and F.~Zuo (AIP
Conference Proceedings 483, 1999), p. 12; cond-mat/9905172. 
\bibitem{Bian} A.~Bianconi, {\it et al.}, {\it Phys.~Rev.~B} {\bf 54},
12018 (1996). 
\bibitem{Eskes}H.~Eskes, {\it et al.}, {\it Phys. Rev. Lett.} {\bf 67},
1035 (1991).
\bibitem{Valla}T.~Valla, {\it et al.}, {\it Science} {\bf 285}, 2110 
(1999); A.~Kaminski, {\it et al.}, {\it Phys. Rev. Lett.} {\bf 84},
1788 (2000).
\bibitem{Comm3}As is mentioned in the text, the stripon states are based
on localized states in charged stripe-like inhomogeneities (such as
shown in Fig.~7). These states are combined to yield the maximal gain in
free energy due to coupling with the QE's and svivons, and combinations
corresponding to areas around points $\pm{\bf k}^p$ in the lattice BZ
are coupled optimally to high-density antinodal QE's through svivons
around their minimum at ${\bf k}_0$. The component ${\pi \over 2a}$ of
${\bf k}^p$ perpendicular to a stripe direction is commensurate with the
distance $4a$ between adjacent charged stripes. Since the charged
stripes sites occupy in this case a quarter of the lattice, the areas
around $\pm{\bf k}^p$, where stripons states reside, cover about a
quarter of the lattice BZ. 
\bibitem{Yoshida1}T.~Yoshida, {\it et al.}, {\it Phys.~Rev.~B} {\bf 63},
220501 (2001). 
\bibitem{Lanzara}A.~Lanzara, {\it et al.}, {\it Nature} {\bf 412}, 510
(2001); X.~J.~Zhou, {\it et al.}, {\it Nature} {\bf 423}, 398 (2003). 
\bibitem{Johnson}P.~D.~Johnson, {\it et al.}, {\it Phys. Rev. Lett.}
{\bf 87}, 177007 (2001). 
\bibitem{Armitage}N.~P.~Armitage, {\it et al.}, , {\it Phys.~Rev.~B}
{\bf 68}, 064517 (2003); cond-mat/0212172. 
\bibitem{Rossat}J.~Rossat-Mignod, {\it et al.}, {\it Physica C} {\bf
185--189}, 86 (1991). 
\bibitem{Fong}H.~F.~Fong, {\it et al.}, {\it Nature} {\bf 398}, 588
(1999); {\it Phys.~Rev.~B} {\bf 61}, 14773 (2000). 
\bibitem{Bourges1}Ph.~Bourges, {\it et al.}, {\it Science} {\bf 288},
1234 (2000); cond-mat/0211227.
\bibitem{Dai}P.~Dai, {\it et al.}, {\it Phys.~Rev.~B} {\bf 63}, 054525
(2001). 
\bibitem{Reznik1}D.~Reznik, {\it et al.}, cond-mat/0307591.
\bibitem{Reznik2}D.~Reznik, {\it et al.}, {\it Phys.~Rev.~B} {\bf 53},
R14741 (1996).
\bibitem{Bourges2}Ph.~Bourges, {\it et al.}, {\it Phys.~Rev.~B} {\bf
56}, R12439 (1997). 
\bibitem{Yoshida2}T.Yoshida, {\it et al.}, {\it Phys. Rev. Lett.}
{\bf 91}, 027001 (2003). 
\bibitem{Comm4}A $t$-hopping of a nearest-neighbor QE pair with
nearest-neighbor spin-flip (see Fig.~6) can be expressed as a combined
process consisting of: ({\it i}) a transition of the QE pair into a pair
state of dressed stripons (at the same sites) by the exchange of a
svivon which turns a pair of nearest-neighbor opposite spins into a
singlet pair of fluctuating spins; ({\it ii}) a $t$-hopping of a
nearest-neighbor stripon pair exchanging their sites with the sites of
the nearest-neighbor fluctuating spins; ({\it iii}) a transition of the
stripon pair back into a QE pair by the exchange of a svivon which turns
the singlet pair of fluctuating spins into nearest-neighbor opposite
spins. 
\bibitem{Comm5}The combination of Eqs.~(25) and (26) results in BCS-like 
equations for both the QE and the stripon order parameters.   
\bibitem{Comm6}$\Phi^q({\bf r}_1,{\bf r}_2)$ does not vanish when ${\bf
r}_1$ and ${\bf r}_2$ are at nearest neighbor QE sites, both in the $a$
and in the $b$ directions, and has sign reversal between the two
directions. Fig.~7 demonstrate this in corner regions, but it is the
case also elsewhere, though there is considerable asymmetry between the
$a$ and the $b$ directions. 
\bibitem{Moca}C.~P.~Moca, and B.~Jank\'o, {\it Phys.~Rev.~B} {\bf 65}, 
052503 (2002); I.~Tifrea, and C.~P.~Moca, cond-mat/0307362.
\bibitem{Emery}V.~J.~Emery, and S.~A.~Kivelson, {\it Nature} {\bf 374},
4347 (1995); {\it Phys.~Rev.~Lett.}~{\bf 74}, 3253 (1995);
cond-mat/9710059. 
\bibitem{Uemura}Y.~J.~Uemura, {\it et al.}, {\it Phys.~Rev.~Lett.} {\bf
62}, 2317 (1989). 
\bibitem{Niedermayer}Ch.~Niedermayer, {\it et al.}, {\it
Phys.~Rev.~Lett.}~{\bf 71}, 1764 (1993). 
\bibitem{marel1}M.~Gr\"uninger, {\it et al.}, {\it Phys.~Rev.~Lett.}
{\bf 84}, 1575 (2000); D.~N.~Basov, {\it Phys.~Rev.~B} {\bf 63}, 134514
(2001); A.~B.~Kuzmenko, {\it et al.}, {\it Phys.~Rev.~Lett.}
{\bf 91}, 037004 (2003).
\bibitem{marel2}H.~J.~A.~Molegraaf, {\it et al.}, {\it Science} {\bf
295}, 2239 (2002); A.~F.~Santander-Syro, {\it et al.}, {\it
Superconducting and Related Oxides: Physics and Nanoengineering V},
edited by I.~Bozovic, and D.~Pavuna (SPIE Proceedings, Volume 4811,
2002), p. 48; {\it Europhys.~Lett.} {\bf 62}, 568 (2003);
cond-mat/0111539; C.~C.~Homes, {\it et al.}, , {\it Phys.~Rev.~B} {\bf
69}, 024514 (2004); cond-mat/0303506. 
\bibitem{Norman}M.~R.~Norman, and C.~Pepin, {\it Phys.~Rev.~B} {\bf 66},
100506 (2002); cond-mat/0302347.
\bibitem{Yusof}Z.~M.~Yusof, {\it et al.}, {\it Phys.~Rev.~Lett.} {\bf
88}, 167006 (2002); A.~Kaminski, {\it et al.}, {\it Phys.~Rev.~Lett.}
{\bf 90}, 207003 (2003).
\bibitem{Feng1}D.~L.~Feng, {\it et al.}, {\it Science} {\bf 289}, 277
(2000); H.~Ding, {\it et al.}, {\it Phys.~Rev.~Lett.} {\bf 87}, 227001
(2001). 
\bibitem{Feng2}D.~L.~Feng, {\it et al.}, {\it Phys.~Rev.~Lett.} {\bf
86}, 5550 (2001); A.~D.~Gromko, {\it et al.}, {\it Phys.~Rev.~B} {\bf
68}, 174520 (2003); cond-mat/0202329; cond-mat/0205385. 
\bibitem{Borisenko}S.~V.~Borisenko, {\it et al.}, {\it Phys.~Rev.~Lett.}
{\bf 90}, 207001 (2003). 
\bibitem{Renner}Ch.~Renner, {\it et al.}, {\it Phys.~Rev.~Lett.} {\bf
80}, 149 (1998); M.~Suzuki, and T.~Watanabe, {\it Phys.~Rev.~Lett.} {\bf
85}, 4787 (2000). 
\bibitem{Kugler}M.~Kugler, {\it et al.}, {\it Phys.~Rev.~Lett.} {\bf
86}, 4911 (2001). 
\bibitem{Zasadzinski}J.~F.~Zasadzinski, {\it et al.}, {\it
Phys.~Rev.~Lett.} {\bf 87}, 067005 (2001). 
\bibitem{Yurgens}A.~Yurgens, {\it et al.}, {\it Phys.~Rev.~Lett.} {\bf
90}, 147005 (2003). 
\bibitem{Peter}B.~W.~Hoogenboom, {\it et al.}, {\it Phys.~Rev.~B} {\bf
67}, 224502 (2003). 
\bibitem{Eschrig}M.~Eschrig, and M.~R.~Norman, {\it Phys.~Rev.~Lett.}
{\bf 89}, 277005 (2002). 
\bibitem{Janowitz}C.~Janowitz, {\it et al.}, cond-mat/0107089.
\bibitem{Pavuna}M.~Abrecht, {\it et al.}, {\it Phys.~Rev.~Lett.} {\bf
91}, 057002 (2003). 
\end{thebibliography}
\end{document}